\documentclass[reprint,
showpacs,
 amsmath,amssymb,
prb,
]{revtex4-1}

\usepackage{graphicx}
\usepackage{dcolumn}
\usepackage{bm}
\usepackage{color}

\begin{document}


\title{Majorana Bound States in Proximity Junctions of Superconducting Nanowires with Dresselhaus Spin-orbit Coupling}
\author{Satoshi Ikegaya$^{1}$}
\author{Yasuhiro Asano$^{1,2,4}$}
\author{Yukio Tanaka$^{3,4}$}%
\affiliation{$^{1}$Department of Applied Physics,
Hokkaido University, Sapporo 060-8628, Japan\\
$^{2}$Center of Topological Science and Technology,
Hokkaido University, Sapporo 060-8628, Japan\\
$^{3}$Department of Applied Physics,Nagoya University, Nagoya 464-8603, Japan\\
$^{4}$Moscow Institute of Physics and Technology, 141700 Dolgoprudny, Russia}%

\date{\today}

\begin{abstract}
We theoretically study transport properties of nanowires 
with the Dresselhaus [110] spin-orbit coupling under the in-plane Zeeman potential
and the proximity-induced $s$-wave pair potential.
In the topologically nontrivial phase, the nanowire hosts the Majorana fermions 
at its edges and the number of the Majorana bound states is equal to
the number propagating channels $(N_c)$. When we attach a normal metal 
to the superconductor, such Majorana bound states penetrate 
into the dirty normal segment and form the $N_c$ resonant transmission channels 
there. We show that chiral symmetry of the electronic states protects
the Majorana bound states at the zero energy even in the presence of impurities. 
 As a result, we find that the zero-bias conductance 
of normal-nanowire / superconducting-nanowire junctions 
is quantized at $2e^{2} N_c/h$ independent of the random potentials.

\end{abstract}

\pacs{Valid PACS appear here}
\maketitle

\section{Introduction}
Majorana Fermion, particle which is own antiparticle, was originally predicted by 
Ettore Majorana in high energy physics\cite{maj1}.
Recently, however, physics of Majorana fermion has been a hot issue in condensed matter physics 
since the emergence of Majorana fermion was pointed out at surfaces of 
topologically nontrivial superconductors\cite{maj2}.
Detecting a Majorana Fermion and controlling of Majorana bound states(MBSs) have been a desired subject 
to realize the fault-tolerant topological quantum computation\cite{tqc1,tqc2}.
There are several suggested systems hosting MBSs such as $p$ wave superconductors\cite{psc1,psc2}, 
topological insulator/superconductor heterostructures\cite{tisw}, 
semiconductor/superconductor junctions with strong spin-orbit interaction\cite{smsc1,smsc2,smsc3,smsc4,smsc5,smsc6},
helical superconductors\cite{hlsc},
and superconducting topological insulators\cite{scti}.
The most practical system among them is a semiconductor nanowire fabricated on top of a superconductor
because of its controllability for the emergence of MBS 
by changing the chemical potential in the nanowire
and by applying the Zeeman field onto it\cite{smsc4,smsc5,smsc6}.
The coexistence of the Rashba spin-orbit coupling and the Zeeman potential enables
a topologically nontrivial superconducting state in the nanowire
in the presence of proximity induced pair potential there.
Even so, it is still very difficult to demonstrate convincing evidences of Majorana fermions in experiments\cite{mjsp1,mjsp2,mjsp3}
because we need to tune the number of propagating channels in nanowires $N_c$.
The $N_c$ should be unity~\cite{smsc4,smsc5} when the Zeeman
field is parallel to the nanowire. 
Alternatively, $N_c$ should be odd integer numbers when the Zeeman field is applied perpendicular 
direction to the nanowire~\cite{mjsp1}. In the latter case, the Zeeman field may destroy the pair potential.

In such situation, we seek an alternative way of realizing Majorana fermion by 
tuning the spin-orbit interactions.
The Dresselhaus spin-orbit interactions are caused by breaking the lattice inversion symmetry\cite{drso}.
In InSb or GaAs, for example, the Dresselhaus [110] spin-orbit
interactions can be large on their film growing along the [110] direction.
A theoretical study has shown that such artificial superconductor hosts 
the dispersionless surface Andreev bound states
which is nothing other than the MBSs\cite{smsc3,flds}. To have topologically nontrivial superconducting state,
the Zeeman field should be applied in plane, which is an advantage of this method.
The [110] Dresselhaus nanowire superconductor is unitary equivalent to
the two-dimensional 'polar state'
 in $^3$He~\cite{leggett}. It has been well known that the
polar state has
surface Andreev bound states~\cite{buchholtz,hara} as a result of the
sign change of the pair potential on the Fermi
surface~\cite{hu,tanaka95}. Today such surface states are recognized as
the topologically protected
edge states reflecting the topologically nontrivial character of the
superconducting phase~\cite{chral}.
Although two of authors have reported the anomalous proximity effect of
superconductors in the
polar state~\cite{yt05r,yt04,ya06,ya07,ya11},
such superconducting state has never been experimentally confirmed in
any compounds.
This paper suggests a way of artificially realizing the 2D polar
superconductor by combining existing materials.
The anomalous proximity effect in the Dresselhaus nanowire
superconducting junctions is strongly related to 
the physics of the odd-frequency Cooper pairs~\cite{dnps4}.

In this paper, we theoretically study the transport properties of nanowires
with strong Dresselhaus [110] spin-orbit interaction
by using the lattice Green function method on the two-dimensional tight-binding lattice.
We first calculate the local density of states (LDOS) at the edge of the semi-infinite nanowire. 
The Dresselhaus [110] nanowire with in-plane magnetic field
shows the large zero-energy peak independent of $N_c$.
The zero-bias differential conductance in normal-metal/superconductor (NS) junctions on the nanowire
shows the quantization at $2e^2N_c/h$ irrespective of the
degree of disorder in the normal segment. 
We also show the fractional current-phase ($J-\varphi$) relationship in 
superconductor/normal-metal/superconductor (SNS) junctions on the nanowire.
The resonant transmission through the MBS in the normal segment is responsible for 
such unusual low energy transport in nanowires\cite{yt05r,yt04,ya06,ya07,dnps1,dnps4}.
In addition to numerical simulation, we solve the Bogoliubov-de-Gennes (BdG) equation analytically
and discuss the stability of MBSs in the the Dresslhaus[110] nanowire with in-plane magnetic fields.
We find that chiral symmetry of the BdG Hamiltonian
protects the MBSs at the zero-energy\cite{chral}.
Our results indicate a way of detecting the MBSs in experiments.

This paper is organized as follows.
In Sec.~II, we compare the local density of states at the edge of the Dresselhaus noanowire 
superconductors with those of the Rashba nanowire superconductors.
The numerical result for the transport properties are also presented.
In Sec.~III, we discuss the stability of the MBSs based on the analytical solution 
of the BdG equation. In Sec.~IV, effects of disorder on the MBS in the normal metal 
are discussed.  
The conclusion is given in Sec.~V.

\section{Numerical Results}
\label{sec:ldos}
\begin{figure}[htbp]
\begin{center}
\includegraphics[width=0.45\textwidth]{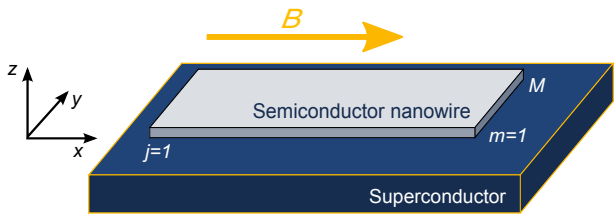}
\caption{(Color online) Schematic picture of a semiconductor nanowire
proximity coupled to a $s$-wave superconductor.}
\label{fig:model1}
\end{center}
\end{figure}
\subsection{Local Density of States}
Let us consider a nanowire with the strong spin-orbit
coupling fabricated on a metallic superconductor as shown in Fig.~\ref{fig:model1}.
The nanowire is in the superconducting state due to the proximity-induced $s$-wave pair potential.
The thickness of nanowire is sufficiently small so that 
only the lowest subband in the $z$ direction for each spin degree of freedom is occupied.
We describe the present nanowire by using the tight-binding model in two-dimension.
A lattice site is pointed by a vector $\boldsymbol{r}=j\boldsymbol{x}+m\boldsymbol{y}$, where $\boldsymbol{x}$ and $\boldsymbol{y}$ 
are the unit vectors in the $x$ and the $y$ directions, respectively.
We consider the nanowire as the semi-infinite system in the $x$ direction (i. e. ,$1 \leq j \leq \infty$).
In the $y$ direction, the number of the lattice site is $M$ and the hard-wall boundary condition is applied.
The nanowire is described by the Bogoliubov-de-Gennes(BdG) Hamiltonian,
\begin{align}
H_{{\rm BdG}} =& H_{{\rm kin}} + H_{Z} + H_{D}^{110} + H_{\Delta}, 
\label{eq:bdg}\\
\hat{H}_{{\rm kin}} =& -t \sum_{\boldsymbol{r},\sigma} \sum_{\boldsymbol{R} =
\boldsymbol{x},\boldsymbol{y}} \left(
c_{\boldsymbol{r} + \boldsymbol{R},\sigma}^{\dagger} c_{\boldsymbol{r} , \sigma}  +  
c_{\boldsymbol{r},\sigma}^{\dagger}c_{\boldsymbol{r} + \boldsymbol{R},\sigma} \right) \nonumber\\
&+ \sum_{\boldsymbol{r},\sigma} ( 4t-\mu ) c_{\boldsymbol{r},\sigma}^{\dagger}c_{\boldsymbol{r},\sigma},
\label{eq:hmkin}
\\
H_{Z} =& - \sum_{\boldsymbol{r},\alpha,\beta} 
V_{ex}(\sigma_{1})_{\alpha,\beta}
c_{\boldsymbol{r},\alpha}^{\dagger} 
c_{\boldsymbol{r},\beta},
\label{eq:hmzm}
\\
H_{D}^{110} =& - i \frac{\lambda_{D}}{2} \sum_{\boldsymbol{r},\alpha,\beta}
(\sigma_{3})_{\alpha,\beta} \left(
c_{\boldsymbol{r} + \boldsymbol{x},\alpha}^{\dagger} c_{\boldsymbol{r} , \beta}  -  
c_{\boldsymbol{r},\alpha}^{\dagger} c_{\boldsymbol{r} + \boldsymbol{x}, \beta} \right),
\label{eq:hmdr}
\\
H_{\Delta} =& \sum_{\boldsymbol{r}} \Delta_{0} \left( 
c_{\boldsymbol{r},\uparrow}^{\dagger} c_{\boldsymbol{r},\downarrow}^{\dagger} + 
H.c. \right),
\end{align}
where $c_{\boldsymbol{r},\sigma}^{\dagger}$($c_{\boldsymbol{r},\sigma}$) is the creation(annihilation) operator of an electron
at the site $\boldsymbol{r}$ with spin $\sigma = ( \uparrow$ or $\downarrow )$,
$t$ denotes the hopping integral, $\mu$ is the chemical potential,
 $\lambda_{D}$ represents the strength of the Dresselhaus [110]  spin-orbit interaction,
$\Delta_{0}$ is the proximity-induced $s$-wave pair potential at the zero temperature.
The Pauli's matrices in spin space are represented by $\hat{\sigma}_{j}$ for $j = 1-3$
and the unit matrix in spin space is $\hat{\sigma}_{0}$.
By tuning the magnetic field $B$ in the $x$ direction as shown in Fig.~\ref{fig:model1}, 
it is possible to introduce the external Zeeman potential $V_{ex}$.
We measure the energy and the length in the units of $t$ and lattice constant, respectively. 
Throughout this paper, we fix several parameters as
 $\mu=1.0t$, $\lambda_{D}=0.2t$ and $\Delta_{0}=0.1t$.
 
At first, we focus on the local density of states (LDOS) at the edge of the nanowire.
The LDOS averaged over $M$ lattice sites in the $y$ direction is defined by
\begin{align}
\rho (j,E) = - \frac{1}{\pi M} \sum_{m} {\rm Im} \left[ {\rm Tr}
\{ \hat{G}(\boldsymbol{r},\boldsymbol{r}, E + i \delta)  \} \right],
\label{eq:ldossm}
\end{align}
where $\hat{G}(\boldsymbol{r},\boldsymbol{r},E)$ is the normal Green's function at the site $\boldsymbol{r}$ with the energy
$E$ measured from the Fermi energy, and ${\rm Tr}$ represents the trace in spin space.
To calculate the LDOS, we add the small imaginary part $i\delta$ to the energy in the Green's function.
We calculate the Green's function by using the lattice Green's function method\cite{grf1,grf2}.
\begin{figure}[htbp]
\begin{tabular}{cccc}
 \begin{minipage}{0.5\hsize}
 \begin{center}
   \includegraphics[width=1.0\textwidth]{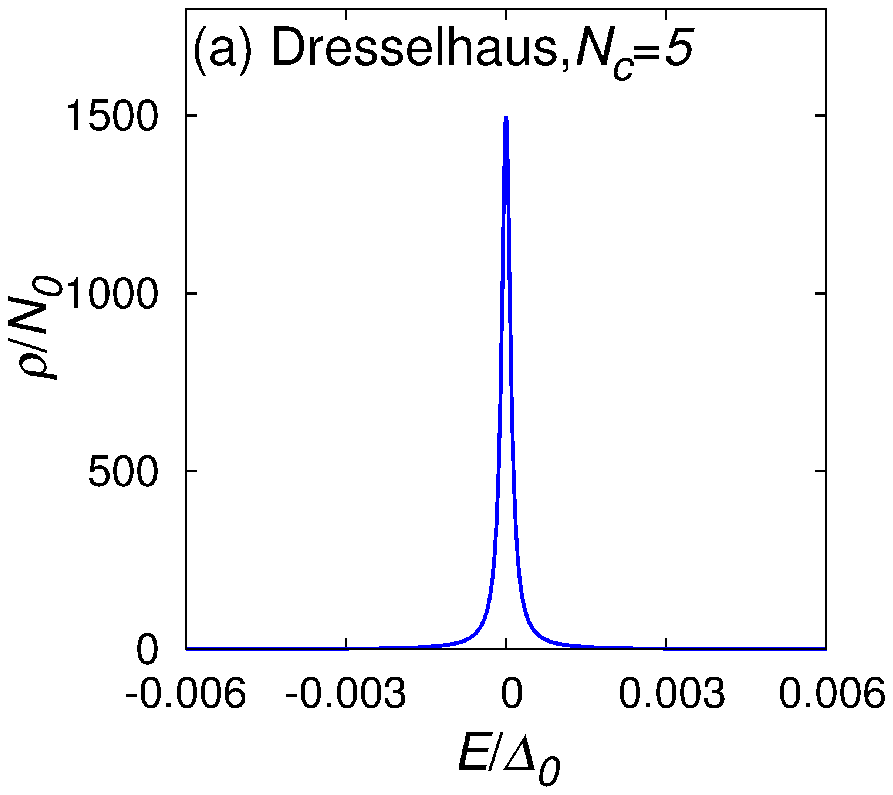}
 \end{center}
 \end{minipage}
 \begin{minipage}{0.5\hsize}
 \begin{center}
   \includegraphics[width=1.0\textwidth]{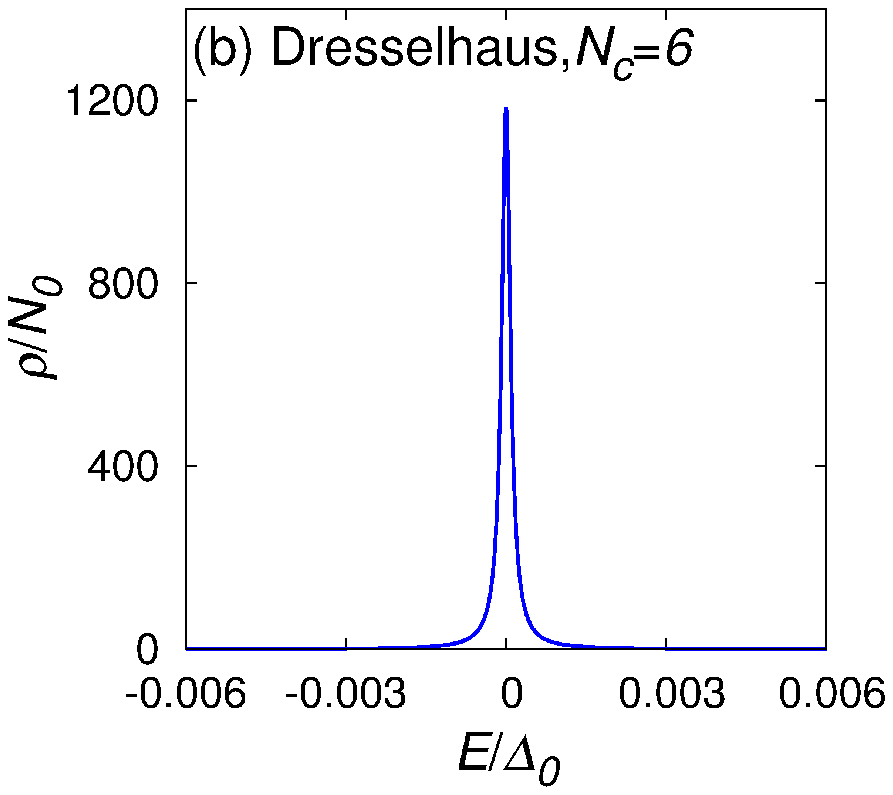}
 \end{center}
 \end{minipage}\\
 \begin{minipage}{0.5\hsize}
 \begin{center}
   \includegraphics[width=1.0\textwidth]{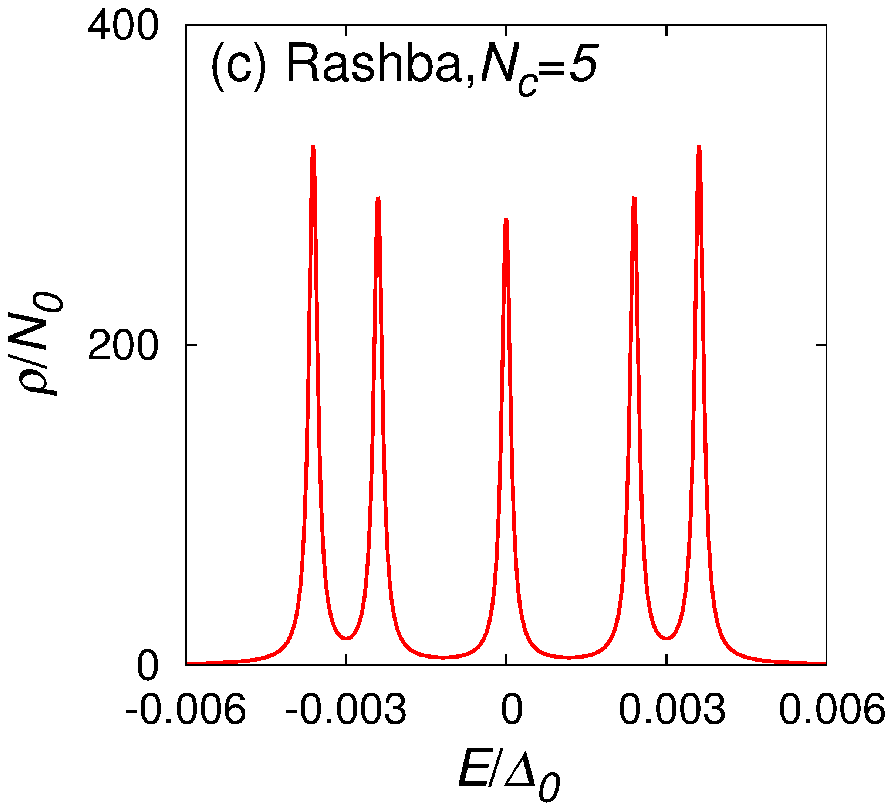}
 \end{center}
 \end{minipage}
 \begin{minipage}{0.5\hsize}
 \begin{center}
   \includegraphics[width=1.0\textwidth]{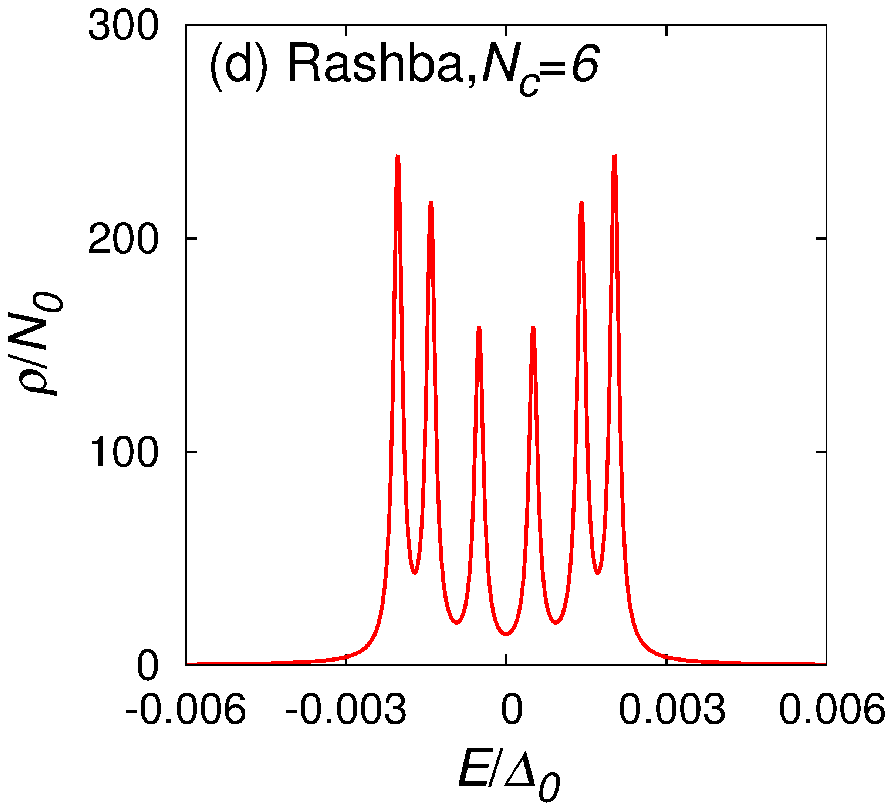}
\end{center}
\end{minipage}
\end{tabular}
 \caption{(Color online)The LDOS at the surface ($j=1$) of superconductors 
 is plotted as a function of the energy.
(a) Dresselhaus [110] nanowire for $V_{ex} = 1.2t$ ($N_{c}=5$).
(b) Dresselhaus [110] nanowire for $V_{ex} = 1.5t$ ($N_{c}=6$).
(c) Rashba nanowire for $V_{ex} = 1.2t$ ($N_{c}=5$).
(d) Rashba nanowire for $V_{ex} = 1.5t$ ($N_{c}=6$).
The samll imaginary part of energy in the Green's function $i \delta$
is chosen as $i 10^{-4} \Delta_{0}$.}
\label{fig:ld-en}
\end{figure}
In Fig.~\ref{fig:ld-en}(a) and (b), we plot the LDOS at $j=1$ as a function of the energy.
The width of system $M$ is chosen as $10$.
The results are normalized by the density of states at the Fermi energy
in the clean normal nanowire $N_{0}$.
The Amplitude of Zeeman potential $V_{ex}$ is
$1.2t$ and $1.5t$ in Fig.~\ref{fig:ld-en}(a) and (b), respectively.
As a result, 
the number of propagating channels $N_{c}$,
is $5$ and $6$ in Fig.~\ref{fig:ld-en}(a) and (b), respectively.
The LDOS at the edge of the Dresselhaus nanowires shows
the single zero-energy peak irrespective of $N_{c}$. 
This is a robust feature
appearing as far as the Zeeman potential $V_{ex}$ is larger than a critical value $V_{c}=0.92t$.
The critical value of the Zeeman potential is discussed in Sec.III C.

For comparison,
we also plots the results for the nanowire with Rashba spin-oribit coupling
in Fig.~\ref{fig:ld-en}(c) and (d). 
To describe the Rashba nanowires, we replace the $H_{D}^{110}$ by
\begin{align}
H_{R} =& - i \frac{\lambda_{R}}{2} \sum_{\alpha,\beta} \sum_{\boldsymbol{r}}
\Bigl[ (\sigma_{2})_{\alpha,\beta} \left(
c_{\boldsymbol{r}c_{\boldsymbol{r} , \beta} + {\bf x},\alpha}^{\dagger} -  
c_{\boldsymbol{r},\alpha}^{\dagger} c_{\boldsymbol{r} + {\bf x} , \beta} \right) \nonumber\\
 &- (\sigma_{1})_{\alpha,\beta} \left(
c_{\boldsymbol{r} + {\bf y},\alpha}^{\dagger} c_{\boldsymbol{r} , \beta}  - 
c_{\boldsymbol{r},\alpha}^{\dagger}c_{\boldsymbol{r} + {\bf y} , \beta} \right) 
\Bigr],
\end{align}
and replace $H_{Z}$ by 
\begin{align}
H_{Z}^{'} = - \sum_{\boldsymbol{r},\alpha,\beta} 
V_{ex}(\sigma_{3})_{\alpha,\beta}
c_{\boldsymbol{r},\alpha}^{\dagger} 
c_{\boldsymbol{r},\beta}.
\end{align}
representing the magnetic field in the $z$ direction.
We chose $V_{ex}$ as $1.2t$ resulting $N_c=5$ in Fig.~\ref{fig:ld-en}(c) and 
$V_{ex}=1.5t$ resulting $N_c=6$ in (d), where $\lambda_{R}=0.2t$ and $M=10$.
As already discussed in Ref.~\onlinecite{mjsp1},
the LDOS of the Rashba nanowire shows the zero-energy peak
only when $N_{c}$ is odd integer numbers.
In addition, the number of peaks in the subgap energy window is equal to $N_{c}$.
Therefore, in the Rashba nanowires, we need the delicate tuning of the wire width and 
the Zeeman field to have the MBS indicated by the zero-energy peak. 

\begin{figure}[tttt]
\begin{tabular}{cc}
 \begin{minipage}{0.25\textwidth}
 \begin{center}
   \includegraphics[width=1.0\textwidth ]{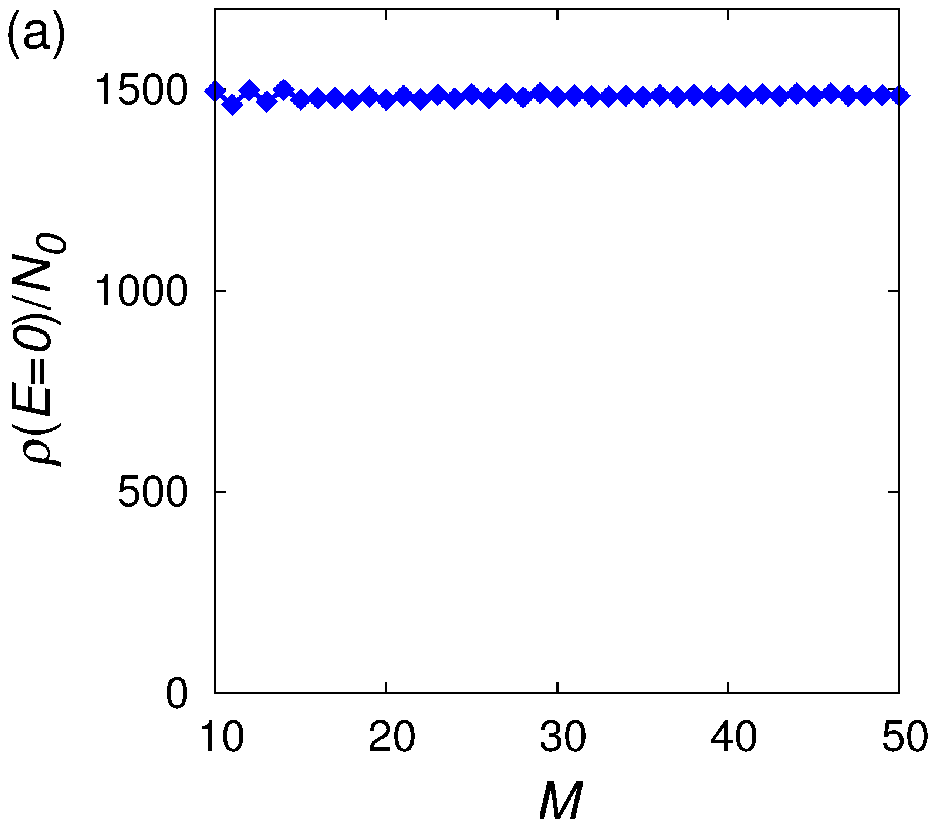}
 \end{center}
 \end{minipage}
 \begin{minipage}{0.25\textwidth}
 \begin{center}
   \includegraphics[width=1.0\textwidth ]{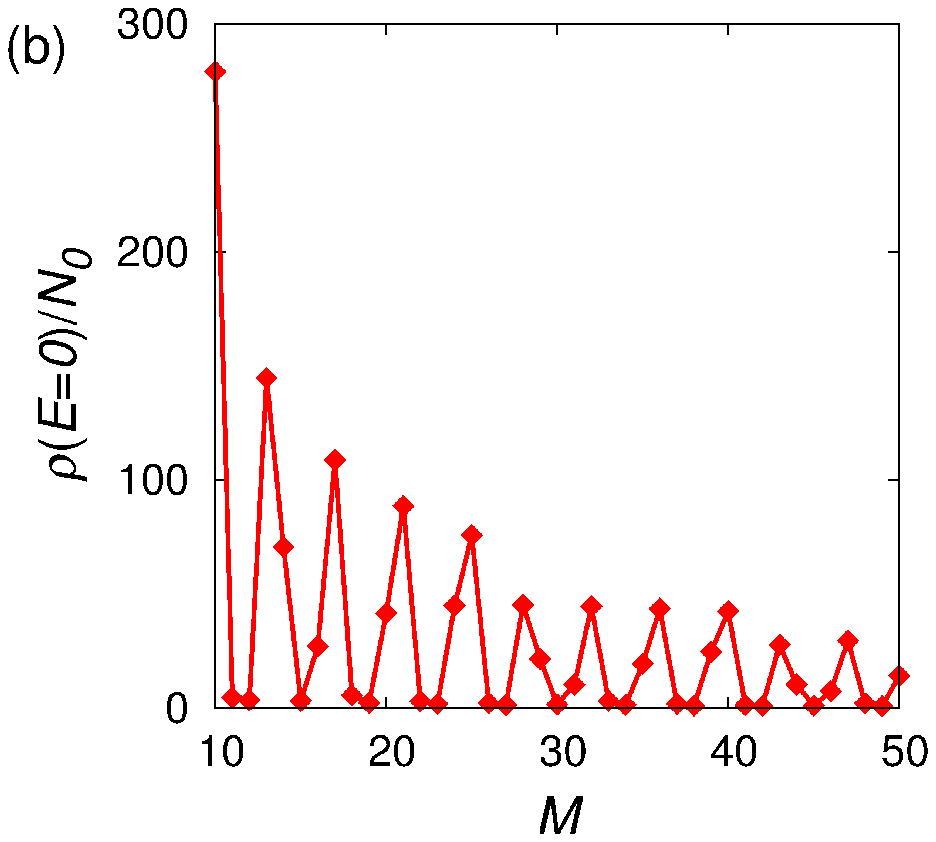}
 \end{center}
 \end{minipage}
\end{tabular}
 \caption{(Color online)The LDOS at the zero-energy is plotted 
 as a function
of the wire width $M$.
The Zeeman potential $V_{ex}$ is chosen as $1.2t$
The results for the Dresselhaus nanowire and for the Rashba nanowire
are plotted in (a) and (b), respectively.
The small imaginary part for the Green's function $i \delta$
is chosen as $i 10^{-4} \Delta_{0}$}
\label{fig:ld-v}
\end{figure}
In Fig.~\ref{fig:ld-v}(a),
we plot the LDOS with $E=0$ at the edge of superconductors
as a function of the wire width $M$ for $V_{ex}=1.2t$.
The LDOS for the Dresselhaus nanowire is almost constant
independent of $M$.
Namely, the number of the zero-energy states at the edge increases proportionally to $M$.
In Fig.~\ref{fig:ld-v} (b),
we also plot the results for the Rashba nanowire.
The LDOS for the Rashba nanowire takes
the zero and the nonzero values alternatively as a function of $M$.
The envelop function of its amplitude
gradually decreases with increasing $M$, which reflects a fact 
that the number of the zero-energy states is at most unity in the Rashba case.

\begin{figure}[hhhh]
\begin{center}
\includegraphics[width=0.48\textwidth]{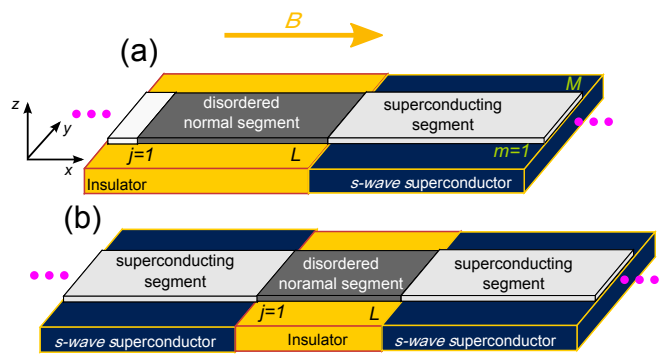}
\caption{(Color online) Schematic pictures of (a) NS and (b) SNS junctions.}
\label{fig:junct}
\end{center}
\end{figure}

\subsection{Conductance}
\begin{figure}[hhhh]
\begin{tabular}{cc}
 \begin{minipage}{0.25\textwidth}
 \begin{center}
   \includegraphics[width=1.0\textwidth]{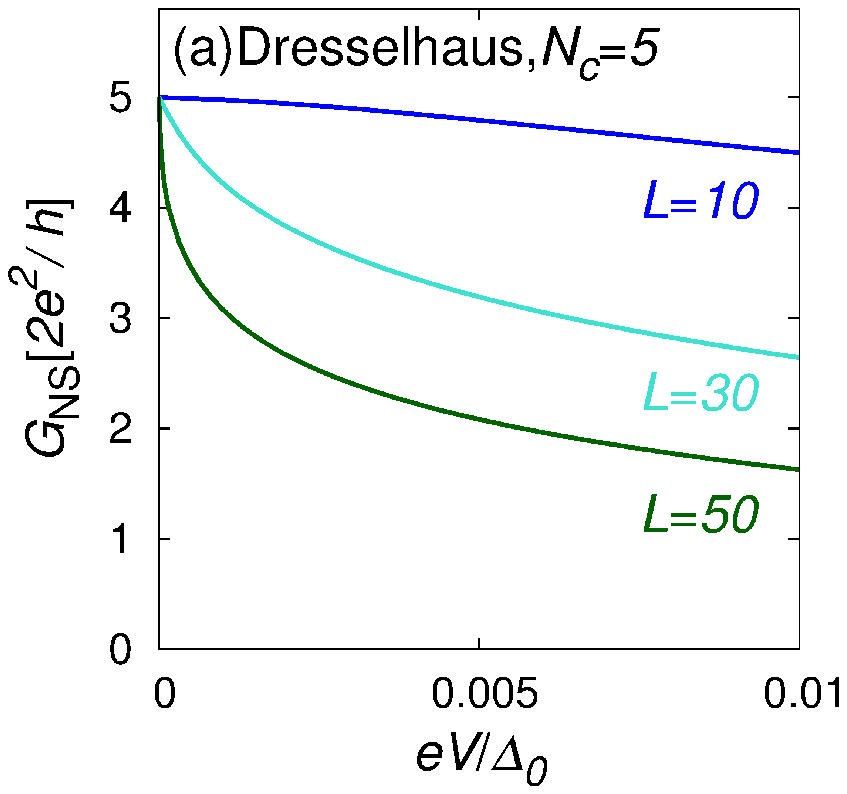}
 \end{center}
 \end{minipage}
 \begin{minipage}{0.25\textwidth}
 \begin{center}
   \includegraphics[width=1.0\textwidth]{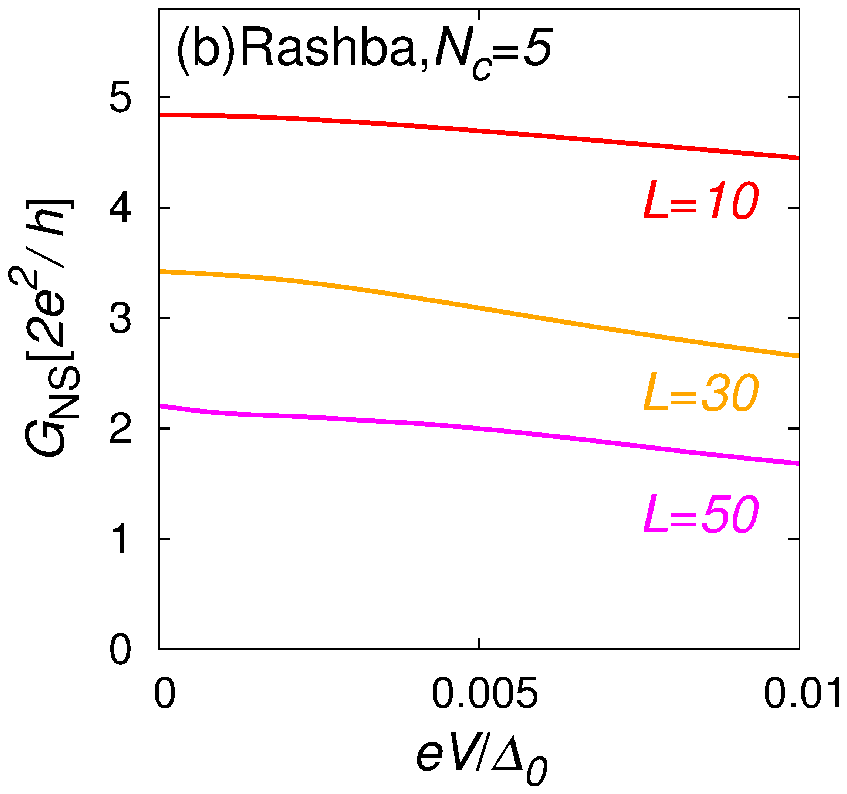}
 \end{center}
 \end{minipage}
\end{tabular}
 \caption{(Color online) The differential conductance is plotted as a function
of the bias voltage for several choices of the length of disordered segment $L$,
where $W=2.0t$, $V_{ex}=1.2t$, and $M=10$.
The results of the Dresselhaus nanowire and those of the Rashba nanowire
are plotted in (a) and (b), respectively.
The number of propagating channels $N_{c}$ is $5$ in both (a) and (b).
The number of samples used for the random ensemble average is $10^{3}.$}
\label{fig:gns-en}
\end{figure}

Secondly we study the conductance in the NS junctions of nanowire superconductors 
as shown in Fig.~\ref{fig:junct}(a).
A nanowire is fabricated on an insulator/metallic superconductor junction.
The segment on the insulator and that on the superconductor are 
in the normal and the superconducting state, respectively.
The present junction consists three segments:
an ideal lead wire ($\infty \leq j \leq 0$), a normal disordered segment ($ 1 \leq j \leq L$)
and a superconducting segment ($L+1 \leq j \leq \infty$).
The superconducting segment is described by
the BdG Hamiltonian in Eq.~(\ref{eq:bdg}).
The normal-metal segment($-\infty \leq j \leq L$) is described by Eq.~(\ref{eq:bdg})
by setting the pair potential $\Delta_{0}$ to zero.
In addition,
we introduce the potential disorder in $1 \leq j \leq L$
by
\begin{gather}
H_{{\rm imp}}=\sum_{\sigma} \sum_{1 \leq j \leq L, m}
V_{0}(\boldsymbol{r}) c_{\boldsymbol{r},\sigma}^{\dagger}c_{\boldsymbol{r},\sigma}.
\label{eq:disoimp}
\end{gather}
The amplitude of impurity potentials is given randomly
in the range of $-W/2 \leq V_{0}(\boldsymbol{r}) \leq W/2$.
We calculate the differential conductance $G_{\textrm{NS}}$
of the NS junctions based on
the Blonder-Tinkham-Klapwijk formula\cite{btkf}
\begin{align}
G_{{\rm NS}}(eV) = \frac{e^{2}}{h} \sum_{\zeta,\eta}
\left[ \delta_{\zeta,\eta} - \left| r^{ee}_{\zeta,\eta} \right|^{2}
+ \left| r^{he}_{\zeta,\eta} \right|^{2} \right]_{eV=E},
\end{align}
where $r^{ee}_{\zeta,\eta}$ and $r^{he}_{\zeta,\eta}$ denote
the normal and Andreev reflection coefficients at the energy $E$, respectively.
The index $\zeta$ and $\eta$ label the outgoing channel and the incoming one, respectively.
These reflection coefficients are calculated
by using the lattice Green's function method\cite{grf1,grf2}.
In Fig.~\ref{fig:gns-en}(a),
we present the differential conductance of the Dresselhaus nanowires
as a function of the bias voltage for several choices of the length of the disordered 
segments $L$, where we choose the parameters as $M=10$, $W=2.0t$, and $V_{ex}=1.2t$.
In the present parameter choice, $N_{c}$ becomes $5$.
The results are the normalized to $G_{Q}=2e^{2}/h$.
The differential conductance decreases with increasing $L$
for the finite bias voltage.
However, the zero-bias conductance is quantized at
$G_{Q}N_{c}$ irrespective of $L$.
The results suggest that the perfect transmission channels exist 
in the disordered normal segment~\cite{yt04} and their number is equal to $N_c$.

For comparison,
we also plot the results for the Rashba nanowire in Fig.~\ref{fig:gns-en}(b).
The results show that the zero-bias conductance is not quantized 
and decrease with increasing $L$.
Even if a MBS appears at the zero-energy, its contribution to the zero-bias conductance 
is relatively small. Therefore it is difficult to demonstrate the presence of MF 
by the conductance measurement in experiments.

\begin{figure}[hhhh]
 \begin{center}
   \includegraphics[width=0.35\textwidth]{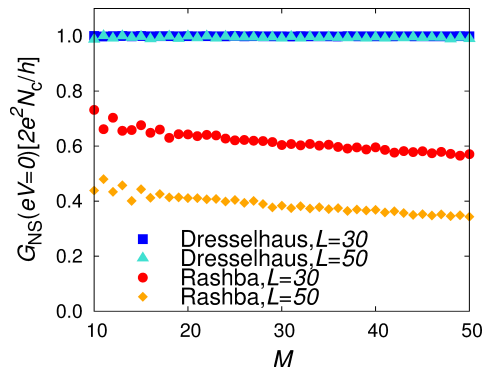}
 \end{center}
 \caption{(Color online) The zero-bias conductance
is plotted as a function of the wire width $M$ for $L=30$ and $50$.
The results are normalized to $G_{Q}N_{c}$.
The number of samples used for the
random ensemble average is $10^{3}.$}
\label{fig:gns-w}
\end{figure}

In Fig.~\ref{fig:gns-w}, we plot the zero-bias conductance
as a function of the width of nanowire $M$ for
$V_{ex}=1.2t$.
The length of disordered segment $L$ is chosen as $30$ and $50$.
In the Dresselhaus nanowires,
the perfect quantization of the zero-bias conductance can be seen
irrespective of the width of the nanowire $M$.
This result reflects the presence of a MBS for each propagating channel.
The presence of MBSs can be checked by the quantized value of the zero-bias conductance. 
%
In the case of the Rashba nanowire, on the other hand, 
the zero-bias conductance slowly decreases with increasing $M$ as shown in Fig.~\ref{fig:gns-w}.
The results are away from the $2e^2N_c/h$ for all $M$.

\subsection{Josephson Current}
\begin{figure}[hhhh]
 \begin{center}
   \includegraphics[width=0.35\textwidth]{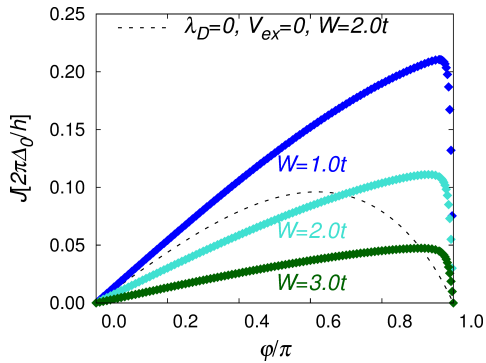}
 \end{center}
 \caption{(Color online) The Josephson current at $T=0.001T_{c}$
is plotted as a function of the phase difference, where $L=30$, $M=10$,
and $V_{ex}=1.5t$.
We plot the result for several choices of $W$
such as $1.0t$, $2.0t$ and $3.0t$.
We also plot the results for the conventional $s$-wave superconductor
(i.e. , $V_{ex}=0$ and $\lambda_{D}=0$) at $W=2.0t$ with a dashed line.
The number of samples used for the
random ensemble average is $500$ for all the plots.}
\label{fig:jcs-p}
\end{figure}

Finally, we study the Josephson effect in the junction shown in Fig.~\ref{fig:junct}(b).
The junction consists three segment: a disordered normal segment ($ 1 \leq j \leq L$)
and two Dresselhaus superconducting nanowires ($\infty \leq j \leq 0$ and $L+1 \leq j \leq \infty$).
The pair potential for the left superconductor is described by
\begin{align}
H_{\Delta}^{L} = \sum_{ j \leq 0, m}  \left(\Delta_{0}e^{i \varphi } 
c_{\boldsymbol{r},\uparrow}^{\dagger} c_{\boldsymbol{r},\downarrow}^{\dagger} + 
H.c. \right),
\end{align}
where $\varphi$ corresponds to the phase difference of the pair potential
between two superconductors.
The Josephson current is calculated by using the Green's function method\cite{jsphs}.
In Fig.~\ref{fig:jcs-p},
we plot the Josephson current at $T=0.001T_{c}$
as a function of the phase difference
for several choices of $W$ such as $1.0t$, $2.0t$ and $3.0t$, where $M=10$, $L=30$, and
$V_{ex}=1.5t$.
For comparison, we also plot the result for the conventional $s$-wave superconductor junction
(i.e., $V_{ex}=0$ and $\lambda_{D}=0$) with a dashed line.
The current-phase relationship for the conventional junction slightly deviate 
the sinusoidal function. The results suggest the small contribution 
of the higher harmonics such as $\sin(2\varphi)$ and $\sin(3\varphi)$ to the Josephson current.
This is the well know behavior of the Josephson current in diffusive SNS junction of the metallic
superconductor\cite{likharev}.
On the other hand,
the Josephson current in the Dresselhaus nanowires indicates the large contribution of the higher
harmonics to the Josephson current at a low temperature.
As a consequence, the results are close to the fractional current-phase relationship 
of $J \propto {\rm sin}(\varphi/2)$ irrespective of $W$.
Such fractional relationship also indicates the perfect transmission through the disordered 
normal segment~\cite{ya06,dnps4}. 
\section{Surface Majorana bound states}
\label{sec:aatmbs}
In this section, we analyze the properties of the Majorana bound states 
appearing at the edge of the Dresslhaus nanowire as shown in Fig.~\ref{fig:model1}.

\subsection{Wave function of zero-energy edge states}
\label{sec:npozees}
Here we consider the nanowire in the continuous space for simplicity.
The BdG Hamiltonian
of the Dresselhaus nanowire is represented by
\begin{align}
\check{H}_{0} = \left[
\begin{array}{cc}
\hat{h} & i \Delta_{0} \hat{\sigma}_{2} \\
- i \Delta_{0} \hat{\sigma}_{2} & - \hat{h}^{*} \\
\end{array}
\right],
\label{eq:original}
\end{align}
\begin{align}
\hat{h} = \xi \hat{\sigma}_{0}
- V_{ex} \hat{\sigma}_{1}
+ i \lambda_D \partial_{x} \hat{\sigma}_{3},\;
\xi = \frac{- \hbar^{2}}{2m} 
\boldsymbol{\nabla}^{2} - \mu, \label{eq:horg1}
\end{align}
where $m$ denotes
the effective mass of an electron.
In what follows, we assume large enough Zeeman potential 
so that $\lambda_D k_F \ll V_{ex}$ is satisfied with $k_F=\sqrt{ 2 m \mu}/\hbar$. 
By applying the unitary transformations in Appendix~\ref{sec:trbdg},
we obtain the deformed BdG Hamiltonian
\begin{align}
\check{H}_{{\rm eff}}
=& \check{H}_{p_{x}} + \check{V}_{\Delta} + O\left[ \left(\frac{\lambda_D k_F}{V_{ex}}\right)^2\right],
\label{eq:bdganap}
\\
\check{H}_{p_{x}}
=& \left[
\begin{array}{cc}
\hat{h}_{p_{x},\uparrow} & 0 \\
0 & \hat{h}_{p_{x},\downarrow} \\
\end{array}
\right],\label{h:pxpm}
\\
\hat{h}_{p_{x}, \sigma}
=& \left[
\begin{array}{cc}
\xi +s_s V_{ex} & -s_s  \frac{\lambda_D \Delta_{0}}{V_{ex}} \partial_{x} \\
s_s \frac{\lambda_D \Delta_{0}}{V_{ex}} \partial_{x} & - \xi -s_s V_{ex} \\
\end{array}
\right], \label{eq:hpxsigma}
\\
\check{V}_{\Delta}
=& \left[
\begin{array}{cc}
0 & i \Delta_{0} \hat{\sigma}_{2} \\
- i \Delta_{0} \hat{\sigma}_{2} &0\\
\end{array}
\right],
\label{eq:spinmix}
\\
s_s=&\left\{ \begin{array}{cl} 
1 & \text{for}\; \sigma=\uparrow \\
-1 & \text{for}\; \sigma=\downarrow.
\end{array}\right.
\end{align}
Since $\lambda_D k_F \ll V_{ex}$,
we ignore the higher order terms indicated by $O[ (\lambda_D k_F/V_{ex})^2]$ in the argument below.
The diagonal components $\hat{h}_{p_{x}, \sigma}$
are equivalent to the Hamiltonian of the spin-triplet $p_{x}$-wave superconductor.
Thus the BdG Hamiltonian represents the spin-full $p_{x}$-wave superconductor $\check{H}_{p_{x}}$
with the spin-mixing term $\check{V}_{\Delta}$.

%

We first solve the BdG equation for the zero-energy edge states by neglecting 
the spin-mixing term $\check{V}_{\Delta}$. We will discuss the effects of 
$\check{V}_{\Delta}$ later on.  
The BdG equation at the zero-energy reads
\begin{align}
\check{H}_{p_{x}} \varphi_{\nu_{0}}(\boldsymbol{r}) = 0.
\label{eq:hbdgzero}
\end{align}
where
\begin{align}
\varphi_{\nu_{0}}(\boldsymbol{r}) \equiv 
\left[ u_{\nu_{0}, \uparrow}(\boldsymbol{r}), v_{\nu_{0}, \uparrow}(\boldsymbol{r}),
u_{\nu_{0}, \downarrow}(\boldsymbol{r}), v_{\nu_{0}, \downarrow}(\boldsymbol{r})
\right]^{\mathrm{T}}
\nonumber
\end{align}
 is the eigen wave function of
the zero-energy states labeled by $\nu_0$.
Under the hard-wall boundary condition in the $y$ direction,
the wave function is represented as
\begin{align}
\varphi_{\nu_{0}}(\boldsymbol{r})
=\sqrt{\frac{2}{M}} \sum_{n} \varphi_{n}(x)
{\rm sin} \left( \frac{n \pi}{M}y \right),
\end{align}
where $n$ indicates the transmission channel in the nanowires and $\varphi_{n}(x)$ 
is a vector with the four components.
In the $x$ direction, we assume that the length of the nanowire is $2L$
 (i.e., $-L \leq x \leq L$)
and we apply the hard-wall boundary conditions at the edge of the nanowire,
\begin{align} 
\varphi_{n}(-L)
= \varphi_{n}(L)
= 0. \label{bcx2}
\end{align}
We show how to solve the BdG equation 
in Appendix~\ref{sec:trbdg}.
Here we summarize the results and discuss important 
properties of the solution. The BdG equation can be solved for each 
transmission channel indicated by $n$ in which 
\begin{align}
\mu_{n} = \mu - \frac{\hbar^{2}}{2m}
\left( \frac{n \pi}{M} \right)^{2} \label{eq:defmun3}
\end{align}
represents the effective chemical potential. 
When $\mu_{n} < - V_{ex}$,
there is no solution of Eq.~(\ref{eq:hbdgzero}) with Eq~(\ref{bcx2}).
For $\mu_{n} > V_{ex}$,
we can find the two solutions for each transmission channel 
at each edge: one is in the spin-up sector and the other is in the spin-down one
in Eq.~(\ref{h:pxpm}). 
At the edge around $x=L$, for example, two zero-energy states in the two spin sectors are degenerate.
However such doubly-degenerate zero-energy states are unstable in the presence of the 
$\check{V}_{\Delta}$ because 
the spin-mixing terms hybridize the two zero-energy states and lift 
the degeneracy. 
Finally, for $ -V_{ex} < \mu_{n} < V_{ex} $,
we obtain the only one zero-energy edge state for each transmission channel at each edge.
Namely the zero-energy state in spin-up sector disappears and 
only the zero-energy state in the spin-down sector remains at each edge.
Since the up-spin state is absent, $\check{V}_{\Delta}$ does not affect such 
zero-energy edge states.
The wave function at the left edge $\varphi_{n}^{L} (x)$ and 
that at the right edge $\varphi_{n}^{R} (x)$
can be represented as
\begin{align}
\varphi_{n}^{L} (x)
=& C_{n}^{L}
\left[
\begin{array}{cccc}
0 \\ 0 \\ 1 \\ -1 \\
\end{array}
\right]
{\rm sin} \left[ \sqrt{k_{n, \downarrow}^{2} - \xi_{D}^{-2}} (x+L) \right]
e^{-x/\xi_{D}}, \label{eq:phil2}
\\
\varphi_{n}^{R} (x)
=& C_{n}^{R}
\left[
\begin{array}{cccc}
0\\ 0 \\ 1 \\ 1 \\
\end{array}
\right]
{\rm sin} \left[ \sqrt{k_{n, \downarrow}^{2} - \xi_{D}^{-2}}  (x-L)\right]
e^{x/\xi_{D}},\label{eq:phir2}
\end{align}
where
\begin{align}
\xi_{D} = \frac{\hbar^{2}V_{ex}}{m \lambda_D \Delta_{0}}, \;
k_{n ,\sigma} = \frac{\sqrt{\mu_{n} -s_s V_{ex}} }{\hbar},
\end{align}
with $C_{n}^{L}$ and $C_{n}^{R}$
being the normalization coefficients.
When $L/\xi_{D} \gg 1$, the two zero-energy states localizing at $x=\pm L$ 
are decoupled from each other.
The number of the zero-energy states at each edge is equal to 
the number of channels which satisfies $ -V_{ex} < \mu_{n} < V_{ex} $.

\subsection{Stability of zero-energy states}
\label{sec:stblityz}
The BdG Hamiltonian in Eq.~(\ref{eq:bdganap})
preserves chiral symmetry,
\begin{align}
\Gamma  \check{H}_{{\rm eff}} \Gamma^{-1} = - \check{H}_{{\rm eff}},
\quad
\Gamma
= \left[
\begin{array}{cc}
\hat{\sigma}_{1} & 0 \\
0 & \hat{\sigma}_{1} \\
\end{array}
\right].
\label{eq:gammsmsec}
\end{align}
In the presence of chiral symmetry,
it is possible to introduce the eigen states of $\check{H}^{2}_{\rm eff}$
\begin{align}
\check{H}^{2}_{\rm eff} \chi_{\lambda}(\boldsymbol{r}) = E^{2}\chi_{\lambda}(\boldsymbol{r}),
\end{align}
where $\chi_{\lambda}(\boldsymbol{r})$ is also the eigen states for $\Gamma$
\begin{align}
\Gamma \chi_{\lambda}(\boldsymbol{r}) = \lambda \chi_{\lambda}(\boldsymbol{r}).
\end{align}
The eigenvalue $\lambda$ is either $+1$ or $-1$.
See also Appendix~\ref{sec:stzs} for details.
By using these eigen states $\chi_{\lambda}(\boldsymbol{r})$,
the states belonging to the zero-energy $\varphi_{\nu_{0}}(\boldsymbol{r})$
can be represented by~\cite{chral}  
\begin{align}
\varphi_{\nu_{0} \lambda}(\boldsymbol{r})
 = \chi_{\nu_{0} \lambda}(\boldsymbol{r}),
\end{align}
where $\chi_{\nu_{0} \lambda}(\boldsymbol{r})$ satisfies
\begin{align}
\check{H}^{2}_{\rm eff} \chi_{\nu_{0} \lambda}(\boldsymbol{r}) = 0.
\end{align}
The index $\nu_0$ labels the eigen states belonging to the zero energy.
From the results above, we conclude that $\varphi_{\nu_{0} \lambda}(\boldsymbol{r})$ 
is the eigen state of $\Gamma$ belonging to $\lambda$, which is an important fact 
leading to the stability of the zero-energy states. 

\begin{figure}[hhhh]
\begin{center}
\includegraphics[width=0.48\textwidth]{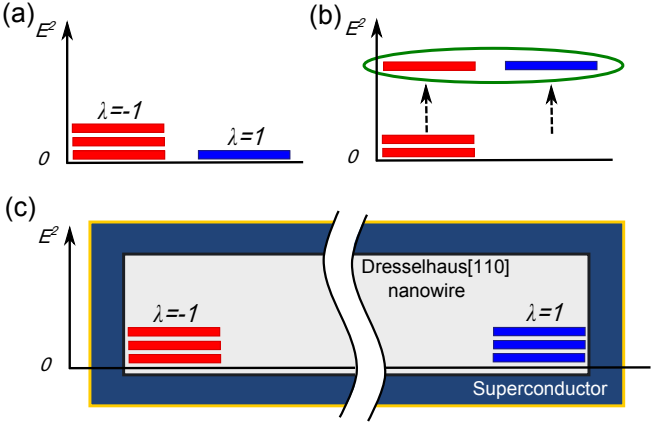}
\caption{(Color online) (a) Four zero-energy states are illustrated.
The three of them belong to $\lambda=-1$ and the one remaining 
state belongs to $\lambda=1$.
(b) Two states with opposite $\lambda$ are coupled with each other
and form the nonzero-energy states.
The remaining two states stay at the zero-energy because
they do not have the coupling partners.
(c) The schematic picture of the zero-energy edge states
of the Dresselhaus nanowire.
The number of the zero-energy states at either edge is equal to $N_c$.
All of the edge states at the left (right) belong to $\lambda=-1$ ($\lambda=1$).
}
\label{fig:sta00}
\end{center}
\end{figure}

The situation in the nonzero-energy states is different that in the zero-energy states.
As shown in Appendix~\ref{sec:stzs},
the nonzero-energy states are always described by the linear combination of two states:
one belongs to $\lambda=1$ (i.e., $\chi_{+}(\boldsymbol{r})$) and
the other belongs to $\lambda=-1$ (i.e., $\chi_{-}(\boldsymbol{r})$).
Generally speaking, perturbations may lift zero-energy states to nonzero-energy ones.
Such modification happens only when the perturbations couple the two zero-energy states 
belonging to opposite $\lambda$ as schematically illustrated in Fig.~\ref{fig:sta00}(a) and (b).
This argument is valid as far as the perturbations preserve chiral symmetry.
Therefore the zero-energy states belong to the same eigenvalue of $\lambda$ 
are stable and remain at the zero-energy when perturbations preserve chiral symmetry
in Eq.~(\ref{eq:gammsmsec}).  
In the Dresselhaus nanowires for $ -V_{ex} < \mu_{n} < V_{ex} $,
it is easy to confirm that $\varphi_n^L$ in Eq.~(\ref{eq:phil2}) belongs to $\lambda=-1$,
whereas $\varphi_n^R$ in Eq.~(\ref{eq:phir2}) belongs to $\lambda=1$.
Since they are spatially separated, the zero-energy states at the two edges are robust 
under perturbations preserving chiral symmetry as illustrated in Fig.~\ref{fig:sta00}(c).

Finally, we note that chiral symmetry in the original basis represented by
\begin{gather}
\check{\Gamma}_{0} \check{H}_0 \check{\Gamma}_0^{-1}= - \check{H}_0,
\\
\Gamma_{0}
= \left[
\begin{array}{cc}
0 & -i \hat{\sigma}_{1}\\
i \hat{\sigma}_{1} & 0 \\
\end{array}
\right], \label{chiral_h0}
\end{gather}
where $\check{H}_0$ is the original Hamiltonian in Eq.~(\ref{eq:original}).
This fact implies that the zero-energy states are robust under perturbations 
preserving chiral symmetry
even if we take the higher order terms of $(\lambda_D k_F/V_{ex})$ into account.

\subsection{Critical value of Zeeman potential}
\label{sec:critzm}
The anomalous properties in the low energy transport in Sec.~\ref{sec:ldos} appear 
when the Zeeman field is larger than a critical value (i.e., $V_{ex}>V_c$). 
Here we discuss why the low energy spectra of the edge states drastically changes 
at $V_{ex}=V_c$. 
We consider the periodic boundary condition in the $x$ direction
to obtain the momentum representation of $\check{H}_{{\rm eff}}$ in Eq~(\ref{eq:bdganap}),
The state vectors can be written as
\begin{align}
\varphi(\boldsymbol{r})
=\sqrt{\frac{2}{LM}} \sum_{n,k} \varphi_{k,n} e^{ikx}
{\rm sin} \left( \frac{n \pi}{M}y \right),
\end{align}
where $n$ labels the channels in the $y$ direction
and $k$ denotes the wave number in the $x$ direction.
\begin{figure}[bbbb]
\begin{tabular}{cc}
 \begin{minipage}{0.23\textwidth}
 \begin{center}
   \includegraphics[width=1.0\textwidth]{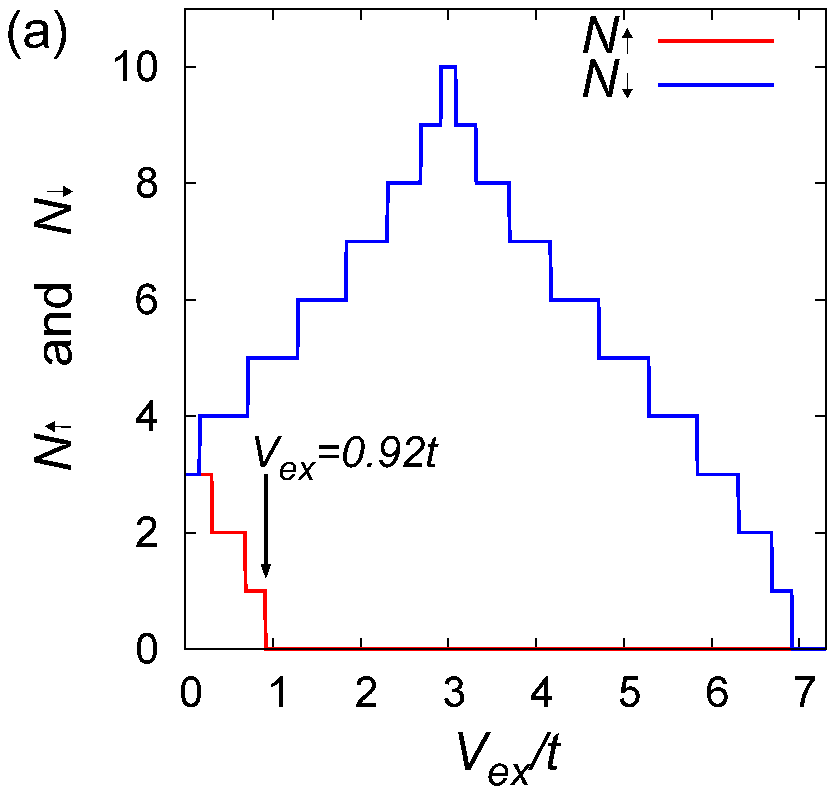}
 \end{center}
 \end{minipage}
 \begin{minipage}{0.23\textwidth}
 \begin{center}
   \includegraphics[width=1.0\textwidth]{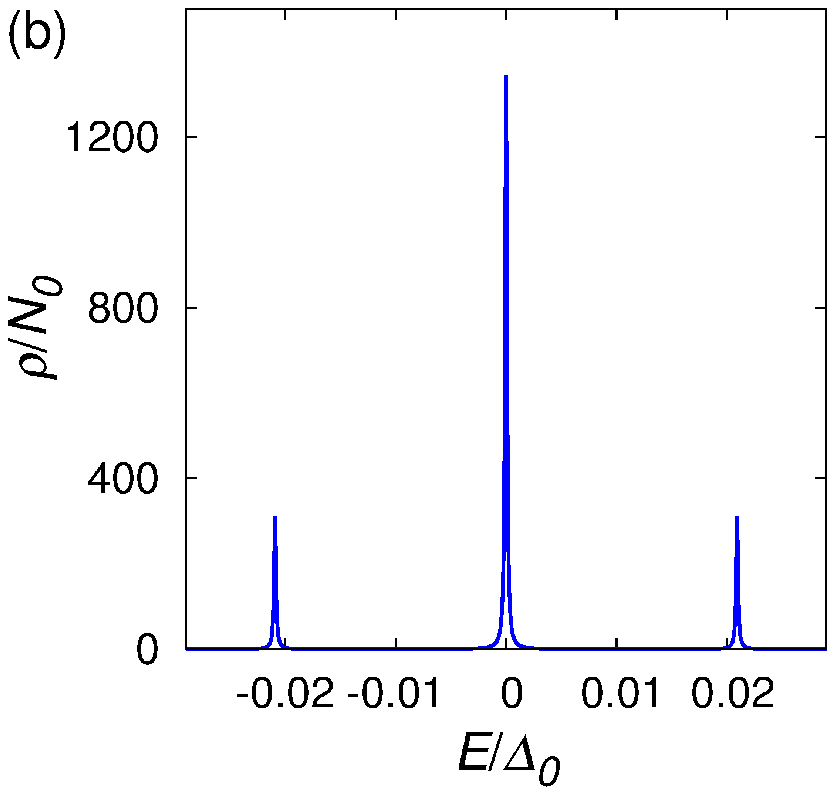}
 \end{center}
 \end{minipage}
\end{tabular}
 \caption{(Color online)(a)
The number of propagating channels $N_{\uparrow}$ and $N_{\downarrow}$ are plotted
as a function of the Zeeman potential $V_{ex}$ at $M=10$.
$N_{\uparrow}$ becomes zero at $V_{ex}=0.92t$.
(b) We plot the local density of states at
the edge of the Dresselhaus nanowire
as a function of energy.
The Zeeman potential $V_{ex}$ is chosen as
$0.8t$ (i.e., $V_{ex} < 0.92t $).
The small imaginary part in the Green's function $i \delta$
is chosen as $i \Delta_{0} \times 10^{-4}$.}
\label{fig:nc-vv}
\end{figure}
The Hamiltonian in Eq.~(\ref{eq:hpxsigma}) in the momentum space is represented by
\begin{align}
\hat{h}_{p_{x}, \sigma}(k,n)
=& \left[
\begin{array}{cc}
\xi_{n}(k) +s_s V_{ex} & -s_s i \frac{\lambda_D \Delta_{0}}{V_{ex}} k \\
s_s i \frac{\lambda_D \Delta_{0}}{V_{ex}} k & - \xi_{n}(k) -s_s V_{ex} \\
\end{array}
\right], \label{eq:px7}
\\
\xi_{n}(k) =&  \frac{\hbar^{2} k^{2}}{2m}  - \mu_{n},
\end{align}
with $\mu_n$ being represented in Eq.~(\ref{eq:defmun3}).
As discussed in Sec.~\ref{sec:stblityz},
the zero-energy edge states are stable 
when the condition
\begin{align}
-V_{ex} < \mu_{n} < V_{ex}
\end{align}
is satisfied.
This condition corresponds to the situation in which the dispersion
of spin-down sector $\xi_{n}(k) - V_{ex}$ remains at the Fermi level and the 
 dispersion of spin-up sector leaves away from the Fermi level (i.e., $\xi_{n}(k) + V_{ex} > 0$).
The number of the zero-energy states becomes equal to the number of the 
propagating channel $N_{c}$ in the spin-down sector.
 In Fig.~\ref{fig:nc-vv}(a),
 we plot the number of the propagating channels
 as a function of the Zeeman potential, where 
$N_{\uparrow(\downarrow)}$ represents the number of the propagating channels
in spin-up (spin-down) sector.
In the tight binding model, 
 the effective chemical potential $\mu_{n}$ should be replaced by
 \begin{align}
 \mu_{n} = \mu - 2t \left[1-
{\rm cos} \left( \frac{n \pi}{M+1} \right) \right].
 \end{align}
 In the spin-up sector, 
  $N_{\uparrow}$ becomes zero at $V_{ex}=V_c=0.92t$ at the present parameter choice.

For $V_{ex}< V_{c}$,
the dispersions in both the spin-up and the spin-down sectors remain at the Fermi level.
In Fig.~\ref{fig:nc-vv}(b), we show the LDOS
at the edge of the Dresselhaus nanowire for $V_{ex}=0.8t < V_c$.
The resulting channel numbers are $N_\uparrow=1$ and $N_\downarrow=5$.
The edge states also arise from the two spin sectors.
But they are interact with each other due to $\check{V}_{\Delta}$.
As a result, the energy of such interacting states leave away from the zero energy.
The results of LDOS show that two peaks appear at $E=\pm 0.021 \Delta_0$ in addition to 
the large zero-energy peak. 
Thus the number of the zero-energy edge states is $N_\downarrow-N_\uparrow$
which is less than the number of propagating channels $N_c=N_\downarrow + N_\uparrow$.

For $V_{ex}>V_c$, the Hamiltonian of the Dresselhous nanowire 
becomes unitary equivalent to that of the two-dimensional spinless 
$p_x$-wave superconductor. It is possible to realize such polar states 
by tuning the spin-orbit coupling in an alternative way as shown in 
Appendix~\ref{sec:eabsoi}. 

\subsection{Majorana Fermions}
\label{sec:maj111}

The field operator of an electron for the BdG Hamiltonian $\check{H}_{{\rm eff}}$
is described as
\begin{gather}
\Psi (\boldsymbol{r}) =
\sum_{\nu}
\left[ \varphi_{\nu}(\boldsymbol{r}) \gamma_{\nu}
+ \Xi \varphi_{\nu}(\boldsymbol{r}) \gamma_{\nu}^{\dagger} \right],\\
\Xi = \Gamma {\cal K},
\end{gather}
where $\gamma_{\nu}^\dagger$ ($\gamma_{\nu}$) is 
the creation (annihilation) operator
of the Bogoliubov quasiparticle belonging to $E_{\nu}$ and
$\Xi$ is the charge conjugation operator with ${\cal K}$ representing the complex conjugation.
Here we focus on the electron operator at the zero-energy states for $V_{ex}>V_c$. 
As discussed in Sec.~\ref{sec:aatmbs} B, the wave function of the zero-energy states are 
well characterized by the channel index $n$ and described by 
\begin{align}
\varphi_{n}^{L}(\boldsymbol{r}) =& \sqrt{\frac{1}{M}}\varphi_n^L(x) {\rm sin}
\left( \frac{n \pi }{M} y \right),\\
\varphi_{n}^{R}(\boldsymbol{r}) =& \sqrt{\frac{1}{M}}\varphi_n^R(x) {\rm sin}
\left( \frac{n \pi }{M} y \right),
\end{align}
where $\varphi_{n}^{L}(\boldsymbol{r})$ ($\varphi_{n}^{R}(\boldsymbol{r})$)
correspond to the left(right) edge states.
These wave function satisfy
\begin{align}
\Gamma \varphi_{n}^{L}(\boldsymbol{r})
=& - \varphi_{n}^{L}(\boldsymbol{r}),
\\
\Gamma \varphi_{n}^{R}(\boldsymbol{r})
=& \varphi_{n}^{R}(\boldsymbol{r}),
\end{align}
as illustrated in Fig.~\ref{fig:sta00}(c).
Therefore, the electron operator of the zero-energy state  
is written as
\begin{align}
\Psi_{n} (\boldsymbol{r}) =&
i\gamma_{n}^{L} (\boldsymbol{r}) + \gamma_{n}^{R} (\boldsymbol{r}),\\
\gamma_{n}^{L} (\boldsymbol{r}) =& -i
\left[
\varphi_{n}^{L}(\boldsymbol{r}) \gamma_{n -}
- \left( \varphi_{n}^{L}(\boldsymbol{r}) \right)^{*} \gamma_{n -}^{\dagger} \right],
\\
\gamma_{n}^{R} (\boldsymbol{r}) =& 
\left[
\varphi_{n}^{R}(\boldsymbol{r}) \gamma_{n +}
+ \left( \varphi_{n}^{R}(\boldsymbol{r}) \right)^{*} \gamma_{n +}^{\dagger} \right],
\end{align}
where the field operator $\gamma_{n}^{L} (\boldsymbol{r})$
and $\gamma_{n}^{R} (\boldsymbol{r})$
correspond to the edge state on the left hand side and that on the right hand side, respectively.
The operator $\gamma_{n}^{L} (\boldsymbol{r})$ 
is pure imaginary 
while $\gamma_{n}^{R} (\boldsymbol{r})$ is real in the present gauge choice. 
It is easy to show that they satisfy the Majorana relation 
\begin{gather}
\left( \gamma_{n}^{L} (\boldsymbol{r}) \right)^{\dagger} = \gamma_{n}^{L} (\boldsymbol{r}),
\label{eq:majo1}
\\
\left( \gamma_{n}^{R} (\boldsymbol{r}) \right)^{\dagger}= \gamma_{n}^{R} (\boldsymbol{r}).
\label{eq:majo2}
\end{gather}
We conclude that 
both $\gamma_{n}^{L}(\boldsymbol{r})$ and $\gamma_{n}^{R}(\boldsymbol{r})$
fields describe the Majorana fermions.
The above relations hold for each propagating channel $n$. Therefore 
the number of Majorana fermions at each edge is $N_\downarrow$ 
which is equal to $N_c$ for $V_{ex}>V_c$.

\section{Majorana bound states in normal metals}
The numerical results in Sec.~II show that the Majorana bound states penetrate into 
the diffusive normal segment and form the resonant transmission channels there.
The perfect transmission through such Majorana bound states at the zero energy 
are responsible for the anomalous transport properties.
Here we discuss why the Majorana bound states remain at the zero energy
 even in the presence of impurity potentials in the normal segment. 

\subsection{Wave function in normal segment}
We first analyze the wave functions in the normal segment using in Eq.~(\ref{eq:px7}).
In the absence of impurity potential, the wave function in the normal segment at $E=0$ 
is decribed by
\begin{align}
\phi_n(x) =& \left[ \begin{array}{c} 1 + r^{ee}_{\uparrow,\uparrow} \\
 r^{he}_{\uparrow,\uparrow} \\ 0 \\ 0 \end{array} \right]
 e^{ ik_{n,\uparrow}x}
 +
 \left[ \begin{array}{c} 0 \\ 0 \\ 1 + r^{ee}_{\downarrow,\downarrow} \\
 r^{he}_{\downarrow,\downarrow}  \end{array} \right]
 e^{ ik_{n,\downarrow}x}, \\
 k_{n,\sigma}=& \sqrt{2m(\mu_n-s_sV_{ex})}/\hbar
 \end{align}
 where $r^{ee}_{\sigma,\sigma}$ is the normal reflection coefficients and 
 $r^{he}_{\sigma,\sigma}$ is the Andreev reflection coefficients. 
Here we consider the wave function 
 at the channel $n$.
The $p_x$ superconductors causes the perfect Andreev reflection at $E=0$, 
which results in $r^{ee}_{\sigma,\sigma} =0$.
At the same time, we obtain $r^{he}_{\uparrow,\uparrow} =1$ and 
$r^{he}_{\downarrow,\downarrow} =-1$. 
We find that the wave function in each spin sector is the eigen state of $\Gamma$ 
in Eq.~(\ref{eq:gammsmsec}). 
Therefore all states in 
spin-up (spin-down) sector belong to $\lambda=1$ ($\lambda=-1$) in the normal segment.
This conclusion is also true even when we introduce impurity potential into the normal segment
because the impurity potentials preserve chiral symmetry in Eq.~(\ref{eq:gammsmsec}).

\subsection{Effects of disorder}
To describe the disordered NS junctions in Fig.~\ref{fig:junct}(a),  
we introduce the random potential $V_{{\rm imp}}(\boldsymbol{r})$ in the normal segment,
(i.e., $0<x<L$).
The Hamiltonian $\hat{h}$ in Eq.~(\ref{eq:original}) is replaced by
\begin{align}
\hat{h}^{{\rm imp}} = \hat{h} + V_{{\rm imp}}(\boldsymbol{r})\hat{\sigma}_{0}.
\end{align}
The Hamiltonian of the NS junction reads,
\begin{align}
\check{H}_{\textrm{NS}} = \left[
\begin{array}{cc}
\hat{h}^{{\rm imp}} & i \Delta_{0} \hat{\sigma}_{2}\Theta(x-L) \\
- i \Delta_{0} \hat{\sigma}_{2}\Theta(x-L) & - [\hat{h}^{{\rm imp}}]^\ast \\
\end{array}
\right].
\label{eq:ns}
\end{align}
It is easy to show that 
 $H_{{\rm NS}}$ satisfies the relations,
\begin{align}
&\Gamma_{0} H_{{\rm NS}} \Gamma_0^{-1} = - H_{{\rm NS}}, \quad 
\Gamma_{0} = \left[ \begin{array}{cc} 0 & -i\hat{\sigma}_1 \\
i\hat{\sigma}_1 & 0 \end{array} \right]. \label{eq:chiral4}
\end{align}
Therefore, all of the zero-energy states at the NS interface 
keep staying at the zero-energy even in the presence of disorder
because all them belong to $\lambda =-1$ as shown in Fig.~\ref{fig:sta00}(c).
Although the channel index $n$ is no longer a good quantum number 
under the potential disorder, the number of the zero-energy states 
is still equal to the number of the propagating channels 
in the spin-down sector $N_\downarrow=N_c$.
This argument can be applied to the zero-energy states
penetrating into the normal segment. 
The resonant transmission channels at the zero energy in the normal 
segment is protected in the presence of chiral symmetry.
This explains the perfect quantization of the zero-bias conductance
at $2e^2N_c/h$. 

\begin{figure}[hhhh]
 \begin{minipage}{0.23\textwidth}
 \begin{center}
   \includegraphics[width=1.0\textwidth]{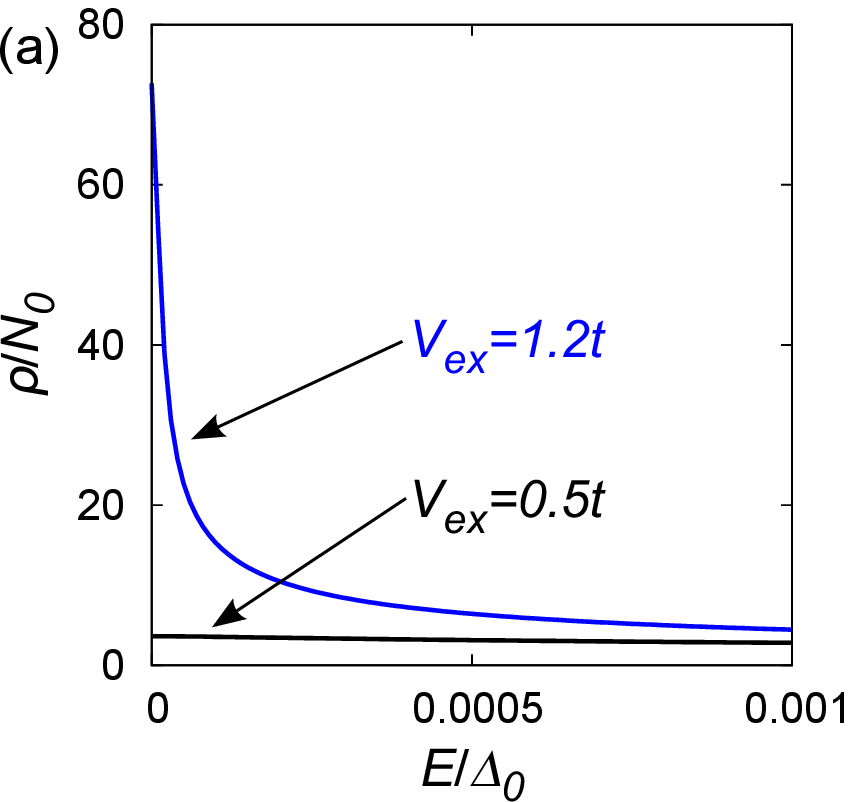}
 \end{center}
 \end{minipage}
 \begin{minipage}{0.23\textwidth}
 \begin{center}
   \includegraphics[width=1.0\textwidth]{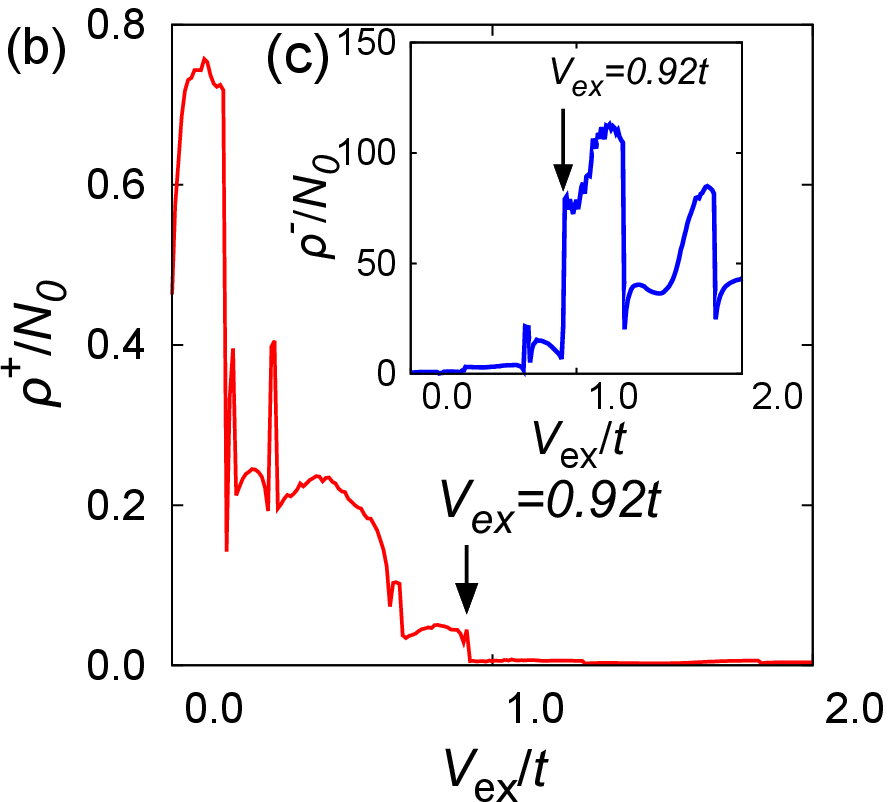}
 \end{center}
 \end{minipage}
 \caption{(Color online)In (a), the LDOS
at the center of disordered segment($j=25$) is plotted
as a function of the energy.
We chose $M=10$, $L=50$ and $W=2.0t$.
We plot the results for $V_{ex}=0.5t$ and $1.2t$.
The LDOS corresponding to  $\lambda=+1$ and
the LDOS corresponding to $\lambda=-1$
is plotted as a function of the Zeeman potential
in (b) and (c), respectively.
The energy for the Green's function is fixed to zero.
We chose $M=10$, $L=50$ and $W=2.0t$ in both (b) and (c).
The number of samples used for
random ensemble average is $10^{4}$ for all the results.
The small imaginary part for the Green's function $i \delta$
is chosen as $i 10^{-5} \Delta_{0}$ for all the results.} 
\label{fig:ns-ld2}
\end{figure}

To confirm the argument above, 
we calculate LDOS in the disordered normal segment.
 In Fig.~\ref{fig:ns-ld2}(a),
we plot the LDOS at the center of the disordered segment (i.e., $j=25$) as a function of the 
energy, where $L=50$, $M=10$, $W=2.0t$.
The Zeeman potential $V_{ex}$ is chosen as $0.5t$ and $1.2t$.
In the case of $V_{ex}<V_c=0.92t$,
the states with $\lambda =1$ remain at the Fermi level in the normal metal.
The random impurity potentials mix
the penetrated Majorana bound states with $\lambda =-1$ and 
the normal states with $\lambda =1$.
As a result,
the LDOS in the disordered normal segment for $V_{ex}=0.5t<V_c$ 
is almost flat around the zero-energy.
When $V_{ex}>V_c$,
all of the normal states with $\lambda =1$ pinch off from the Fermi level.
Therefore,
penetrated Majorana bound states stably remain at the Fermi level.
As a result,
the LDOS in the disordered normal segment for $V_{ex}=1.2t>V_c$
shows the large zero-energy peak
reflecting the existence of the penetrated Majorana bound states.
It is possible to demonstrate how chiral symmetry protects the zero-energy states 
in the normal segment by analyzing the details of LDOS at the zero-energy in Fig.~\ref{fig:ns-ld2}(a). 
To do this, we calculate the Green's function around the zero energy.
The Hamiltonian satisfies,
\begin{align}
&\Xi_{0} H_{{\rm NS}} \Xi_0^{-1} = - H_{{\rm NS}},
\quad
\Xi_{0} = \left[ \begin{array}{cc} 0 & \mathcal{K}\hat{\sigma}_0 \\
\mathcal{K}\hat{\sigma}_0 & 0 \end{array} \right], \label{eq:charge4}
\end{align}
where $\Xi_{0}$ represents the charge conjugation.
When $\phi_{{\nu}}(\boldsymbol{r}) \equiv 
\left[ {u}_{{\nu},\uparrow}(\boldsymbol{r}), {u}_{{\nu},\downarrow}(\boldsymbol{r}),
{v}_{{\nu},\uparrow}(\boldsymbol{r}), {v}_{{\nu},\downarrow}(\boldsymbol{r})
\right]^{\mathrm{T}}$ is the wave function belonging to $E_{{\nu}}$, 
$\bar{\phi}_{{\nu}}(\boldsymbol{r})=\Xi_0 \phi_{{\nu}}(\boldsymbol{r})$ 
is the wave function belonging to $-E_{{\nu}}$.
Using these wave functions, the Green's function is represented by
\begin{align}
\check{G}(\boldsymbol{r},\boldsymbol{r};E)
=\sum_{{\nu}}\left[
 \frac{ \phi_\nu(\boldsymbol{r}) \phi^\dagger_\nu(\boldsymbol{r}') }{E+i\delta - E_\nu }
+
 \frac{ \bar{\phi}_{{\nu}}(\boldsymbol{r}) \bar{\phi}^\dagger_{{\nu}}(\boldsymbol{r}')}{E+i\delta + E_\nu}
\right].
\end{align}
When we consider $|E|\ll \Delta_0$, the wave functions at the zero energy mainly contribute to the 
Green's function,
\begin{align}
\check{G}(\boldsymbol{r},\boldsymbol{r};E)
\approx \sum_{\nu_0} \frac{\left[
 \phi_{\nu_0}(\boldsymbol{r}) \phi^\dagger_{\nu_0}(\boldsymbol{r}') 
+
\bar{\phi}_{\nu_0}(\boldsymbol{r}) \bar{\phi}^\dagger_{\nu_0}(\boldsymbol{r}')
\right]}{E+i\delta},
\end{align}
where $\nu_0$ indicates the states at the zero energy.
We can immediately confirm that
\begin{align}
[ H_{NS}^{2},\Gamma_0 ]=[ H_{NS}^{2},\Xi_0 ]=[ \Gamma_0, \Xi_0 ] = 0.
\end{align}
Therefore, as shown in Appendix~A,
we find that $\phi_{\nu_0}(\boldsymbol{r})$ is the eigen vector of $\Xi_0$
and $\Gamma_0$ at the same time. 
>From this fact, it is possible to represent the wave function by the linear combination of 
following vectors
\begin{align}
\Phi_{\lambda,\alpha,\nu_0}(\boldsymbol{r})
=
\frac{1}{2}\left[\begin{array}{c} a_{\lambda,\alpha,\nu_0}(\boldsymbol{r})
 \\ -\lambda \;\alpha \; a_{\lambda,\alpha,\nu_0}^\ast(\boldsymbol{r}) \\
 -i  \alpha \; a_{\lambda,\alpha,\nu_0}^\ast(\boldsymbol{r})\\ 
 i \lambda  \; a_{\lambda,\alpha,\nu_0}(\boldsymbol{r}) \end{array} \right], 
\end{align} 
where $\Phi_{\lambda,\alpha,\nu_0}$ is the eigen function of 
$\Gamma_0$ belonging to $\lambda=\pm 1$
and is also the eigen function of $\Xi_0$ belonging $i\alpha$ with $\alpha=\pm 1$. 
We assume $\int d\boldsymbol{r} |{a}_{\lambda,\alpha,\nu_0}(\boldsymbol{r})|^2=1$.
By using these vectors, the normal Green's function at the electron space becomes
\begin{align}
&\hat{G}(\boldsymbol{r},\boldsymbol{r}';E) = \hat{G}^+(\boldsymbol{r},\boldsymbol{r}';E) + \hat{G}^-(\boldsymbol{r},\boldsymbol{r}';E),\\
&\hat{G}^\pm(\boldsymbol{r},\boldsymbol{r}';E) = \frac{1}{4(E+i\delta)}\sum_{\alpha,\nu_0} \nonumber\\
\times  &
 \left[ \begin{array}{cc} 
 a_{\pm,\alpha,\nu_0}(\boldsymbol{r})a_{\pm,\alpha,\nu_0}^\ast(\boldsymbol{r}') 
 & \mp \alpha \;a_{\pm,\alpha,\nu_0}(\boldsymbol{r})a_{\pm,\alpha,\nu_0}(\boldsymbol{r}')\\
\mp \alpha \; a_{\pm,\alpha,\nu_0}^\ast(\boldsymbol{r})a_{\pm,\alpha,\nu_0}^\ast(\boldsymbol{r}')
&
 a_{\pm,\alpha,\nu_0}^\ast(\boldsymbol{r})a_{\pm,\alpha,\nu_0}(\boldsymbol{r}') 
 \end{array} \right]. \nonumber
\end{align}
The normal Green's function
$\hat{G}^+$ ($\hat{G}^-$) are the Green function derived from the wave function with $\lambda=1$ ($\lambda=-1$).
In the same way, the anomalous Green function is represented as 
\begin{align}
&\hat{F}(\boldsymbol{r},\boldsymbol{r}';E) = \hat{F}^+(\boldsymbol{r},\boldsymbol{r}';E) + \hat{F}^-(\boldsymbol{r},\boldsymbol{r}';E),\\
&\hat{F}^\pm(\boldsymbol{r},\boldsymbol{r}';E) = \frac{i}{4(E+i\delta)}  \sum_{\alpha,\nu_0}\nonumber\\
\times &
 \left[ \begin{array}{cc} 
 \alpha\; a_{\pm,\alpha,\nu_0}(\boldsymbol{r})a_{\pm,\alpha,\nu_0}(\boldsymbol{r}') 
 & \mp  a_{\pm,\alpha,\nu_0}(\boldsymbol{r})a_{\pm,\alpha,\nu_0}^\ast(\boldsymbol{r}')\\
\mp a_{\pm,\alpha,\nu_0}^\ast(\boldsymbol{r})a_{\pm,\alpha,\nu_0}(\boldsymbol{r}')
& \alpha\; a_{\pm,\alpha,\nu_0}^\ast(\boldsymbol{r})a_{\pm,\alpha,\nu_0}^\ast(\boldsymbol{r}') 
 \end{array} \right]. \nonumber
\end{align}
By using the relations,
\begin{align}
\hat{F}^\pm (\boldsymbol{r},\boldsymbol{r};E) \; \hat{\sigma}_1=& \mp i \hat{G}^\pm(\boldsymbol{r},\boldsymbol{r};E),
\end{align}
$\hat{G}^\pm$ is described by $\hat{G}$ and $\hat{F}$ obtained in the numerical simulation,
\begin{align}
\hat{G}^\pm (\boldsymbol{r},\boldsymbol{r};E) 
=\frac{1}{2}\left[ \hat{G}(\boldsymbol{r},\boldsymbol{r};E) \pm i \hat{F}(\boldsymbol{r},\boldsymbol{r};E) \hat{\sigma}_1
\right]. \label{eq:rofag}
\end{align}
We can separate the numerical results of LDOS at the zero-energy into two contributions
\begin{gather}
\rho (j,E)=\rho^{+}(j,E) + \rho^{-}(j,E),
\\
\rho^{\pm}(j,E) = - \frac{1}{\pi M} \sum_{m} {\rm Im} \left[ {\rm Tr}
\{ \hat{G}^{\pm}(\boldsymbol{r},\boldsymbol{r};E)  \} \right].
\end{gather}
In Fig.~\ref{fig:ns-ld2}(b) and (c),
we plot the LDOS corresponding to the states with $\lambda=1$ (i.e., $\rho^{+}(j,0)$) and
with $\lambda=-1$(i.e., $\rho^{-}(j,0)$)
as a function of the Zeeman potential $V_{ex}$, respectively.
Here the LDOS are calculated at the center (i.e., $j=25$) of the disordered segment with $L=50$.
When $V_{ex}< V_c = 0.92 t$,
the penetrated Majorana bound states with $\lambda=-1$ and 
the normal states with $\lambda=1$ are coupled by the random disordered potentials.
As a result, the amplitude of the LDOS at the zero energy remains at a small value.
As shown in Fig.~\ref{fig:ns-ld2}(b),
$\rho^{+}(j,0)$
become zero for $V_{ex} > V_c$, which means the all zero-energy states belong to $\lambda=1$ 
disappear.
As shown in Fig~\ref{fig:ns-ld2}(c),
$\rho^{-}(j,0)$ suddenly becomes large for $V_{ex} > V_c$.
All of the penetrated Majorana bound states belong to $\lambda=-1$.
As discussed in Sec.~IIIC,
the number of the zero-energy states at the edge of superconductor is equal to
the number of the propagating channels $N_{c}$ for $V_{ex} > V_c$.
In the normal segment, the number of the zero-energy states also becomes $N_c$ even in the presence of 
disorder because the impurity potentials preserve chiral symmetry. 
Such Majorana bound states at the zero energy form the resonant transmission channels in the normal segment.
This explains the perfect transmission in the presence of disorder.

\subsection{Odd-frequency Cooper pairs}
In the normal segment in Fig.~4(a), the zero energy states 
consist of two contributions: one is the penetrated 
MBSs from the superconductor and the other is the usual 
metallic states at the Fermi level. 
The LDOS in Fig.~9(a) indicates that the former is much 
dominant than the latter for $V_{ex}>V_c$.
According to the argument in Sec.~III B, such MBSs in the normal segment 
belong to $\lambda=-1$. 
Therefore the anomalous Green's function can be represented by $F^{-}$. 
By applying the analytic continuation, we obtain
\begin{align}
&\hat{\mathcal{F}}^-(\boldsymbol{r},\boldsymbol{r}';i\omega_n) = \frac{1}{4\omega_n}  \sum_{\alpha,\nu_0}\nonumber\\
\times &
 \left[ \begin{array}{cc} 
 \alpha\; a_{-,\alpha,\nu_0}(\boldsymbol{r})a_{-,\alpha,\nu_0}(\boldsymbol{r}') 
 &   a_{-,\alpha,\nu_0}(\boldsymbol{r})a_{-,\alpha,\nu_0}^\ast(\boldsymbol{r}')\\
 a_{-,\alpha,\nu_0}^\ast(\boldsymbol{r})a_{-,\alpha,\nu_0}(\boldsymbol{r}')
& \alpha\; a_{-,\alpha,\nu_0}^\ast(\boldsymbol{r})a_{-,\alpha,\nu_0}^\ast(\boldsymbol{r}') 
 \end{array} \right]. \nonumber
\end{align}
The anomalous function satisfies
\begin{align}
\hat{\mathcal{F}}^-(\boldsymbol{r},\boldsymbol{r}';i\omega_n) = 
\left[\hat{\mathcal{F}}^-(\boldsymbol{r}',\boldsymbol{r};i\omega_n)\right]^{\mathrm{T}}.
\end{align}
Thus the pairing correlation is spin-triplet even-parity.
As a result, the pairing correlation is the odd function of $\omega_n$.
Therefore Majorana fermions always accompany the odd-frequency Cooper pairs~\cite{dnps4}. 
The odd-frequency Cooper 
pairs support the quantization of the zero-bias conductance in Sec.~II B~\cite{yt04,ya07}. 
The Josephson current shown in Sec.~II C is carried by the odd-frequency Cooper 
pairs~\cite{ya06}.

\section{Conclusion}
We have studied transport properties in junctions consisting of a superconducting 
nanowire with Dresselhaus[110] spin-orbit coupling.
The local density of states at the edge of the isolated nanowire shows the large 
zero-energy peak when the Zeeman potential is larger than a critical value.
 This single peak structure reflects the existence of the Majorana bound states. 
We show that the number of such Majorana bound states 
is equal to the number of the propagating channels $N_c$. 
When we attach a normal nanowire to the superconducting one,
the Majorana bound states penetrate into the normal segment and 
form $N_c$ resonant transmission channels there. 
All of the Majorana bound states remains at the zero energy 
because of chiral symmetry of the junction.
As a result, the Majorana bound states in the normal segment
are responsible for the perfect transmission.
We numerically show that the zero-bias differential conductance 
of the normal-metal/superconductor junction
are quantized at $2e^{2} N_{c}/h$ irrespective of the disorder.
The Josephson current in disordered superconductor/normal-metal/superconductor 
junctions shows the fractional current-phase relationship
$J \propto \sin (\varphi/2)$ at a low temperature.
The superconducting nanowires with Dresselhaus[110] spin-orbit coupling
are two-dimensional analog of the spin-triplet superconductor in the 'polar' state.
Our results indicate a way of detecting Majorana Fermions in experiments.

\begin{acknowledgments}
The authors are grateful to J. D. Sau for useful discussion.
This work was partially supported by the ``Topological Quantum
Phenomena'' (Nos.~22103002, 22103005)
Grant-in Aid for
Scientific Research on Innovative Areas and KAKENHI (No.~26287069) from
the Ministry of Education,
Culture, Sports, Science and Technology (MEXT) of Japan
and by the Ministry of Education and Science of the Russian Federation
(Grant No.~14Y.26.31.0007).
\end{acknowledgments}

\appendix
\section{Zero energy states under chiral symmetry}
\label{sec:stzs}
Here, we briefly summarize the argument in Ref.~\onlinecite{chral}
which shows the important properties
of the zero-energy states under the chiral symmetry.
We consider the BdG Hamiltonian $H$ which preserves the chiral symmetry
\begin{align}
\Gamma H \Gamma^{-1} = - H, \;
\Gamma^{2}=1,
\label{eq:c-cs}
\end{align}
where Eq.~(\ref{eq:c-cs}) is equivalent to
\begin{align}
[ H^2, \Gamma ] = 0.
\label{eq:c-cs2}
\end{align}
The BdG equation is given by
\begin{align}
H \varphi_E (\boldsymbol{r}) = E \varphi_E (\boldsymbol{r}).
\label{eq:c-be}
\end{align}
When we consider the eigen equation
\begin{align}
H^2 \chi_{E^{2}}(\boldsymbol{r}) = E^{2} \chi_{E^{2}}(\boldsymbol{r}),
\label{eq:chidefa}
\end{align}
Eq. (\ref{eq:c-cs2}) suggest that the states $\chi_{E^{2}} (\boldsymbol{r})$ 
is also the eigen states of $\Gamma$ at the same time.
Since $\Gamma^{2} = 1$,
we find that the eigen value of $\Gamma$ is $+1$ or $-1$.
Namely the eigen equation
\begin{align}
 \Gamma \chi_{E^2\lambda}(\boldsymbol{r})
= \lambda \chi_{E^2 \lambda}(\boldsymbol{r}),
\label{eq:c-gc}
\end{align}
holds for $\lambda = \pm 1$.
By multiplying $H$ to Eq.~(\ref{eq:c-gc})
from the left side
and by using Eq.~(\ref{eq:c-cs}),
we obtain the equation
\begin{align}
\Gamma H \chi_{E^{2} \lambda}(\boldsymbol{r})
= - \lambda H \chi_{E^{2} \lambda}(\boldsymbol{r}).
\end{align}
We find that $H \chi_{E^{2} \lambda}(\boldsymbol{r})$ is the eigen state of $\Gamma$ 
belonging to $-\lambda$.
Thus we can connect $\chi_{E^{2} +}(\boldsymbol{r})$ and $\chi_{E^{2} -}(\boldsymbol{r})$
as
\begin{align}
H \chi_{E^{2} \lambda}(\boldsymbol{r}) = 
c_{E^{2} \lambda} \chi_{E^{2} -\lambda}(\boldsymbol{r}),
\label{eq:c-chr}
\end{align}
where $c_{E^{2} \lambda}$ is a constant.
As shown in Ref.~\onlinecite{chral},
the one-to-one correspondence exists between 
$\varphi_E (\boldsymbol{r})$ and
$\chi_{E^2}(\boldsymbol{r})$.

At first, we consider zero energy states 
$\chi_{0 \lambda}(\boldsymbol{r})$ which satisfies
\begin{align}
H^2 \chi_{0 \lambda}(\boldsymbol{r}) = 0,
\end{align}
in Eq.~(\ref{eq:chidefa}). The integration of $\boldsymbol{r}$ after 
multiplying $\chi_{0 \lambda}^\dagger(\boldsymbol{r})$ from the left
results in
\begin{align}
\int d\boldsymbol{r} \left| H \chi_{0 \lambda}(\boldsymbol{r}) \right|^{2} = 0.
\end{align}
This means that the norm of $H \chi_{0 \lambda}(\boldsymbol{r})$ is zero.
Therefore we conclude that 
\begin{align}
H \chi_{0 \lambda}(\boldsymbol{r})=0. \label{eq:hchizero}
\end{align}
As a result, we find the relation
\begin{align}
\varphi_{0 \lambda}(\boldsymbol{r})
 = \chi_{0 \lambda}(\boldsymbol{r}).
\label{eq:c-pec}
\end{align}
When a zero energy state is described by $\varphi_{0 +}(\boldsymbol{r})
 = \chi_{0 +}(\boldsymbol{r})$, the relations in Eqs.~(\ref{eq:c-chr}) and (\ref{eq:hchizero})
  suggest that $\chi_{0 -}(\boldsymbol{r})=0$.
Therefore the zero-energy states are always the eigen states of $\Gamma$.

For $E \neq 0$, it is possible to represent $\varphi_{E}(\boldsymbol{r})$ 
by $\chi_{E^2 \pm}(\boldsymbol{r})$~\cite{chral}.
By calculating the norm of $H \chi_{E^{2} \lambda}(\boldsymbol{r})$,
we obtain
\begin{align}
E^{2}=\left| c_{E^{2} \lambda} \right|^{2}.
\end{align}
Multiplying $H$ to Eq. (\ref{eq:c-chr}) from the left alternatively
gives a relation
\begin{align}
c_{E^{2} \lambda}c_{E^{2} -\lambda} = 1.
\end{align}
Therefore,
we find the relation
\begin{align}
H \chi_{E^{2} \lambda}(\boldsymbol{r}) = 
E e^{ i\lambda \theta_{E^{2}}} \chi_{E^{2} -\lambda}(\boldsymbol{r}).
\label{eq:c-chr2}
\end{align}
Although we cannot fix the phase factor $\theta_{E^{2}}$, 
it is possible to express the states $\varphi_E (\boldsymbol{r})$ for $E \neq 0$
as
\begin{align}
\varphi_E (\boldsymbol{r}) =& \frac{1}{\sqrt{2}}
\left( e^{-i\theta_{E^{2}} /2} \chi_{E^{2} +}(\boldsymbol{r})
+s_E  e^{i\theta_{E^{2}} /2}\chi_{E^{2} -}(\boldsymbol{r}) \right),
\label{eq:epm2}
\\
s_E=&\left\{ \begin{array}{cl} 
1 & \text{for}\; E>0 \\
-1 & \text{for}\; E<0.
\end{array}\right.
\end{align}
The nonzero-energy states are constructed
by a pair of eigen states for $\Gamma$: one belongs to $\lambda=1$ and the other belongs $\lambda=-1$.
Therefore, the states with $E \neq 0$ are not the eigen states of $\Gamma$.
On the contrary to the nonzero-energy states,
the the zero-energy states are the eigen states of $\Gamma$.

\section{Description of zero-energy edge states}
\label{sec:trbdg}
The BdG Hamiltonian
of the Dresselhaus nanowire is represented by $\check{H}_0$ in Eq.~(\ref{eq:original}).
By using the unitary matrix
\begin{align}
\check{R} = \left[
\begin{array}{cc}
\hat{r} & 0 \\
0 & \hat{r}^{*} \\
\end{array}
\right],\;
\hat{r}= \frac{1}{\sqrt{2}}
\left[
\begin{array}{cc}
e^{-i \pi/4} & -e^{-i \pi/4} \\
e^{i \pi/4} & e^{i \pi/4} \\
\end{array}
\right],
\label{eq:unir}
\end{align}
the BdG Hamiltonian $\check{H}_0$ is first transformed to
\begin{align}
\check{H}^{\prime} 
&= \check{R} \check{H}_{0} \check{R}^{\dagger}
\nonumber\\ 
&= \left[
\begin{array}{cc}
\hat{h}^{\prime} & i \Delta_{0} \hat{\sigma}_{2} \\
- i \Delta_{0} \hat{\sigma}_{0} & - \hat{h}^\prime \\
\end{array}
\right],
\label{eq:bdg1}\\
\hat{h}^{\prime} &= \xi \hat{\sigma}_{0}
+ V_{ex} \hat{\sigma}_{3}
+ i \lambda_D \partial_{x} \hat{\sigma}_{2}.
\end{align}
The Hamiltonian in this basis is represented only by real numbers.
Next we apply a transformation
which is similar to the Foldy-Wouthysen transformation~\cite{fwts}
to the BdG Hamiltonian in Eq.~(\ref{eq:bdg1}). 
Using a unitary matrix
\begin{align}
\check{U} =& \left[
\begin{array}{cc}
\hat{u} & 0 \\
0 & \hat{u} \\
\end{array}
\right],
\label{eq:uniu}\\
\hat{u} =&{\rm exp} [ i \hat{S} ],\;
\hat{S} = \frac{\lambda_D}{2 \hbar V_{ex}} p_{x} \hat{\sigma}_{1},
\end{align}
with $p_{x}= -i \hbar \partial_{x}$,
we transform $H'$ into 
\begin{align}
\check{U} \check{H}^{'} \check{U}^{\dagger}
= \left[
\begin{array}{cc}
e^{i\hat{S}} \hat{h}^{'} e^{-i\hat{S}} & e^{i\hat{S}} (i \Delta_{0} \hat{\sigma}_{2} ) e^{-i\hat{S}}\\
- e^{i\hat{S}} (i \Delta_{0} \hat{\sigma}_{2} ) e^{-i\hat{S}} & - e^{i\hat{S}} \hat{h}^{'} e^{-i\hat{S}} \\
\end{array}
\right].
\end{align}
The diagonal term of Eq.~(\ref{eq:bdg1}) can be expanded as
\begin{align}
e^{i\hat{S}} \hat{h}^{\prime} e^{i\hat{S}} 
= \hat{h}^{\prime} + i [ \hat{S}, \hat{h}^{'} ]
+\frac{i^{2}}{2 !} [ \hat{S}, [ \hat{S}, \hat{h}^{\prime} ] ]
+ \cdots,
\end{align}
with using the Baker-Housdorff formula.
We assume large enough Zeeman potential so that $\lambda_D k_{F} \ll V_{ex}$ is satisfied where
$k_{F} = \sqrt{2m \mu}/\hbar$ denotes Fermi wave number.
>From this assumption, we obtain
\begin{align}
e^{i\hat{S}} \hat{h} e^{i\hat{S}}
=\xi \hat{\sigma}_{0}
+ V_{ex} \hat{\sigma}_{3} + O\left[\left(\frac{\lambda_D k_{F}}{V_{ex}}\right)^2\right],
\end{align}
within the first order of $\lambda_D k_F/ V_{ex}$.
The off-diagonal term corresponding to the pair potential
is transformed to
\begin{align}
e^{i\hat{S}} (i \Delta_{0} \hat{\sigma}_{2} ) e^{-i\hat{S}}
&= i \Delta_{0} \hat{\sigma}_{2} + i [ \hat{S}, i \Delta_{0} \hat{\sigma}_{2} ] + \cdots
\nonumber\\
&= i \Delta_{0} \hat{\sigma}_{2} - i \frac{\lambda_D \Delta_{0}}{\hbar V_{ex}} p_{x} \hat{\sigma}_{3}
+ O\left(\frac{(\lambda_D k_{F})^{2}}{V_{ex}^{2}}\right),
\end{align}
where we assume the uniform pair potential (i.e., $[p_{x},\Delta_{0}]=0$).
As the result, the BdG Hamiltonian can be written as
\begin{align}
\check{U} \check{H}^{\prime} \check{U}^{\dagger}
= &\left[
\begin{array}{cccc}
\xi + V_{ex} & 0 & - i \frac{\lambda_D \Delta_{0}}{\hbar V_{ex}} p_{x} & \Delta_{0} \\
0 & \xi - V_{ex} & - \Delta_{0} &  i \frac{\lambda_D \Delta_{0}}{\hbar V_{ex}} p_{x} \\
i \frac{\lambda_D \Delta_{0}}{\hbar V_{ex}} p_{x} & -\Delta_{0} & -\xi - V_{ex} & 0 \\
\Delta_{0} &  - i \frac{\lambda_D \Delta_{0}}{\hbar V_{ex}} p_{x} & 0 & -\xi + V \\
\end{array}
\right] \nonumber\\
&+ O\left(\frac{(\lambda_D k_{F})^{2}}{V_{ex}^{2}}\right).
\end{align}
By interchanging the second column and the third one,
and by interchanging the second row and the third one,
the Hamiltonian can be deformed as
\begin{align}
\check{H}_{{\rm eff}}
=& \check{H}_{p_{x}} + \check{V}_{\Delta},
\label{eq:dfbdg}
\\
\check{H}_{p_{x}}
=& \left[
\begin{array}{cc}
\hat{h}_{p_{x},\uparrow} & 0 \\
0 & \hat{h}_{p_{x},\downarrow} \\
\end{array}
\right],
\\
\hat{h}_{p_{x}, \sigma}
=& \left[
\begin{array}{cc}
\xi + s_s V_{ex} & - s_s i \frac{\lambda_D \Delta_{0}}{\hbar V_{ex}} p_{x} \\
s_s i \frac{\lambda_D \Delta_{0}}{\hbar V_{ex}} p_{x} & - \xi -s_s V_{ex} \\
\end{array}
\right],
\\
\check{V}_{\Delta}
=& \left[
\begin{array}{cc}
0 & i \Delta_{0} \hat{\sigma}_{2} \\
- i \Delta_{0} \hat{\sigma}_{2} &0\\
\end{array}
\right],\\
s_s=&\left\{ \begin{array}{cl} 
1 & \text{for}\; \sigma=\uparrow \\
-1 & \text{for}\; \sigma=\downarrow.
\end{array}\right.,
\end{align}
where the diagonal components $\hat{h}_{p_{x}, \sigma}$
are equivalent to the Hamiltonian of the spin-triplet $p_{x}$-wave superconductor.
Therefore, Eq.~(\ref{eq:dfbdg}) corresponds to
the BdG Hamiltonian of the spin-full $p_{x}$-wave superconductor $\check{H}_{p_{x}}$
with the spin-mixing term $\check{V}_{\Delta}$.
In addition, we find that the BdG Hamiltonian $\check{H}_{{\rm eff}}$ preserves chiral symmetry
\begin{align}
\Gamma  \check{H}_{{\rm eff}} \Gamma^{-1} = - \check{H}_{{\rm eff}},
\quad
\Gamma
= \left[
\begin{array}{cc}
\hat{\sigma}_{1} & 0 \\
0 & \hat{\sigma}_{1} \\
\end{array}
\right].
\end{align}
\par
Now we seek the wave function of the zero-energy edge states
in $\check{H}_{{\rm eff}}$.
We first neglect the spin-mixing term $\check{V}_{\Delta}$.
After solving
the BdG equation at the zero-energy with $\check{V}_{\Delta}=0$
\begin{align}
\check{H}_{p_{x}} \varphi_{\nu_{0}}(\boldsymbol{r}) = 0,
\label{eq:bdgeq0}
\end{align}
we will discuss effects of $\check{V}_{\Delta}$.
As shown in Appendix~\ref{sec:stzs},
the zero-energy states under chiral symmetry
is also the eigen states of $\Gamma$
\begin{align}
 \Gamma \varphi_{\nu_{0} \lambda}(\boldsymbol{r})
= \lambda \varphi_{\nu_{0} \lambda}(\boldsymbol{r}),
\label{eq:add-ch}
\end{align}
where $\lambda = \pm 1$.  
>From Eq.~(\ref{eq:add-ch}), 
the zero-energy states can be written as
\begin{align}
\varphi_{\nu_{0}\lambda}(\boldsymbol{r}) =
[ u_{\nu_{0} \lambda, \uparrow}(\boldsymbol{r}),
\lambda u_{\nu_{0} \lambda, \uparrow}(\boldsymbol{r}),
u_{\nu_{0} \lambda, \downarrow}(\boldsymbol{r}),
\lambda u_{\nu_{0} \lambda, \downarrow}(\boldsymbol{r})
]^{\mathrm{T}}.
\label{eq:spi0}
\end{align}
By substituting Eq.~(\ref{eq:spi0}) into the BdG equation,
we obtain
\begin{align}
\hat{h}_{p_{x} ,\sigma}
\left[
\begin{array}{cc}
u_{\nu_{0} \lambda, \sigma}(\boldsymbol{r}) \\
\lambda u_{\nu_{0} \lambda, \sigma}(\boldsymbol{r})\\
\end{array}
\right] = 0,
\end{align}
which is equivalent to
\begin{align}
\left[ \xi +s_s V_{ex}  - s_s i \lambda \frac{\lambda_D \Delta_{0}}{\hbar V_{ex}} p_{x}
\right] u_{\nu_{0} \lambda, \sigma}(\boldsymbol{r}) = 0.
\label{eq:eqspin}
\end{align}
When we apply the hard-wall boundary condition in the 
$y$ direction, the wave function can be expanded as
\begin{align}
\varphi_{\nu_{0} \lambda, \sigma}(\boldsymbol{r})
=\sqrt{\frac{2}{M}} \sum_{n=1}^M u_{n, \lambda, \sigma} (x)
{\rm sin} \left( \frac{n \pi}{M}y \right),
\label{eq:spin}
\end{align}
where ${M}$ denotes the width of the nanowire.
By substituting Eq. (\ref{eq:spin}) into Eq. (\ref{eq:eqspin}),
we obtain the equations for each transmission channel $n$,
\begin{align}
\left[ \partial_{x}^{2} - s_s 2 \frac{\lambda}{\xi_{D}}\partial_{x} + k_{n, \sigma}^{2} \right]
u_{n, \lambda, \sigma}(x) = 0,
\label{eq:upspin}
\end{align}
where
\begin{gather}
\xi_{D} = \frac{\hbar^{2}V_{ex}}{m \lambda_D \Delta_{0}},
\\
k_{n ,\sigma} = \frac{\sqrt{\mu_{n} -s_s V_{ex}} }{\hbar},\;
\mu_{n} = \mu - \frac{\hbar^{2}}{2m}
\left( \frac{n \pi}{M} \right)^{2}. \label{defmun}
\end{gather}

In what follows, we consider the nanowire with the length $2L$ in the $x$ direction (i.e., $-L \leq x \leq L$)
and apply the hard-wall boundary condition at the edge of the nanowire,
\begin{align} 
u_{n, \lambda, \sigma}(-L)
= u_{n, \lambda, \sigma}(L)
= 0.
\label{eq:bc2}
\end{align}
The length of the nanowire is long enough so that $L/\xi_{D} \gg 1$ is satisfied.
For $\mu_{n} < V_{ex}$,
there is no solution which satisfies
the boundary conditions in Eq.~(\ref{eq:bc2}).

For $ -V_{ex} < \mu_{n} < V_{ex} $,
we find the two solutions in spin-down sector as
\begin{align}
\varphi_{n,\downarrow}^{L} (x)
=& \frac{C_{L}}{\sqrt{2}}
\left[
\begin{array}{cccc}
0 \\ 0 \\ 1 \\ -1 \\
\end{array}
\right]
{\rm sin} \left[ \sqrt{k_{n, \downarrow}^{2} - \xi_{D}^{-2}} (x+L) \right]
e^{-x/\xi_{D}},\label{eq:leftedg}
\\
\varphi_{n,\downarrow}^{R} (x)
=& \frac{C_{R}}{\sqrt{2}}
\left[
\begin{array}{cccc}
0\\ 0 \\ 1 \\ 1 \\
\end{array}
\right]
{\rm sin} \left[ \sqrt{k_{n, \downarrow}^{2} - \xi_{D}^{-2}}  (x-L)\right]
e^{x/\xi_{D}},
\label{eq:rightedg}
\end{align}
where $C_{L}$ and 
 $C_{R}$ are the normalization coefficients.
It is easy to show that $\varphi_{n}^{L} (x)$ localizing at the left edge belongs to $\lambda=-1$ 
and $\varphi_{n}^{R} (x)$ localizing at the right edge
belongs to $\lambda =1$.
%
In the spin-up sector, on the other hand, there is no solution.

For $\mu_{n} > V_{ex}$,
there are solutions at either edges in both spin-up and spin-down sectors.
For the left edge,
we find
\begin{align}
\varphi_{n,\uparrow}^L(x)
= \frac{C_{L}^{'}}{2}
\left[
\begin{array}{c}
1 \\ 1 \\ 0 \\ 0 \\
\end{array}
\right]
 {\rm sin} \left[ \sqrt{k_{n, \uparrow}^{2} - \xi_{D}^{-2}} (x+L) \right]
e^{-x/\xi_{D}},
\label{eq:leftedg2}
\end{align}
in addition to $\varphi_{n,\downarrow}^L(x)$ in Eq. (\ref{eq:leftedg}).
The solution $\varphi_{n,\uparrow}^L(x)$ belongs to $\lambda=1$.
The two zero-energy states at the left edge (i.e., $\varphi_{n,\uparrow}^L(x)$ and $\varphi_{n,\downarrow}^L(x)$)
belong to the opposite spin sector and to the opposite $\lambda$ to each other.
At the right edge, we find
\begin{align}
\varphi_{n,\uparrow}^R(x)
= \frac{C_{R}^{'}}{2}
\left[
\begin{array}{c}
1 \\ -1 \\ 0 \\ 0 \\
\end{array}
\right]
 {\rm sin} \left[ \sqrt{k_{n, \uparrow}^{2} - \xi_{D}^{-2}} (x-L) \right]
e^{x/\xi_{D}}
\end{align}
in addition to $\varphi_{n,\downarrow}^R(x)$ in Eq.~(\ref{eq:rightedg}).
The two zero-energy states at the right edge (i.e., $\varphi_{n,\uparrow}^R(x)$ and $\varphi_{n,\downarrow}^R(x)$ )
belong to the opposite spin sector and to the opposite $\lambda$ to each other.

Next, we consider the effects of
the spin-mixing term $\check{V}_{\Delta}$ in Eq.~(\ref{eq:dfbdg}).
For $\mu_{n} > V_{ex}$,
the two zero-energy states exist for each propagating channel at each edge. 
At the left edge, for example, the spin-up state with $\lambda=1$ 
and the spin-down state with $\lambda=-1$ coexist.
The spin-mixing term $\check{V}_{\Delta}$ couples the two states.
As a result, the energy of the coupled states shifts away from the zero-energy 
because the linear combination of a state with $\lambda=1$ and a state with $\lambda=-1$ 
can form the nonzero-energy state as discussed in Eq.~(\ref{eq:epm2}).
In this case, all the edge states can leave away from the zero-energy.
For $ -V_{ex} < \mu_{n} < V_{ex} $, on the other hand,
 the only one zero-energy state with spin-down exists for each propagating channel at each edge
because the zero-energy state with spin-up is absent.
The edge states remain at the zero-energy even in the presence of $\check{V}_{\Delta}$
because coupling partners are absent.
In other words, $\check{V}_{\Delta}$
does not affect the zero-energy edge states.

\section{Effective $p_{x}$-wave superconductor with coexistence
of Rashba and Dresselhaus[100] spin-orbit coupling}
\label{sec:eabsoi}
\begin{figure}[htbp]
\begin{center}
\includegraphics[width=0.48\textwidth]{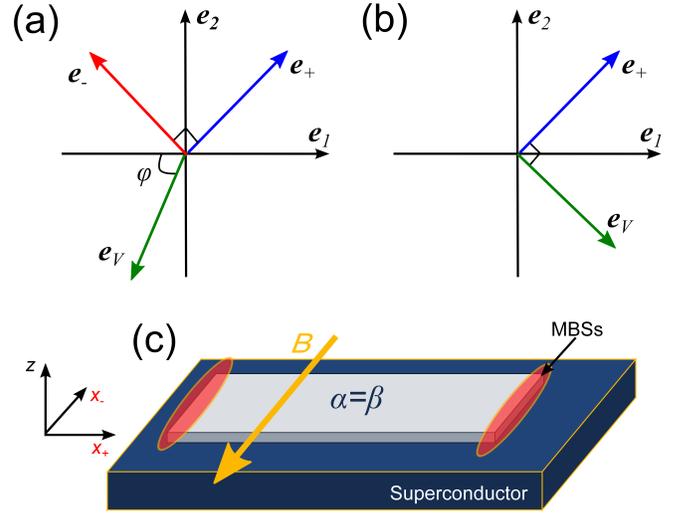}
\caption{(Color online) (a) 
The directions $\boldsymbol{e}_+$ and $\boldsymbol{e}_-$ are orthogonal to each other 
in the spin-space.
The direction of $\boldsymbol{V}$ is denoted by $\boldsymbol{e}_V$.
(b) At $\alpha = \beta'$,
the spin-orbit interaction proportional to $\hat{\sigma}_{-}$ vanishes.
When we chose the Zeeman potential as $V_{x}=-V_{y}$,
$\boldsymbol{e}_{+}$ becomes
perpendicular to $\boldsymbol{e}_{V}$.
(c) Schematic picture of the effective $p_{x}$-wave superconductor
with the coexistence of the Rahba and Dresselhaus [100]
spin-orbit coupling.}
\label{fig:appc}
\end{center}
\end{figure}

We propose an alternative nanowire whose Hamiltonian is 
unitary equivalent to Eq.~(\ref{eq:original}). 
Let us consider the two dimensional electron system
where the Rashba spin-orbit coupling and the Dresselhaus[100]
one coexist.
The Hamiltonian is represented by
\begin{align}
&\hat{h}_{RD}=\hat{h}_{{\rm kin}} + \hat{h}_{R} + \hat{h}_{D}^{100} + \hat{h}_{V}, \\
&\hat{h}_{{\rm kin}} =\frac{1}{2m} \left( p_{x}^{2} + p_{y}^{2} \right)
 - \mu, \\
&\hat{h}_{R} = \frac{\alpha}{\hbar}\left( p_{y} \hat{\sigma}_{1} - p_{x} \hat{\sigma}_{2} \right), \\
&\hat{h}_{D}^{100} = \frac{\beta}{\hbar}\left( p_{x} \hat{\sigma}_{1} - p_{y} \hat{\sigma}_{2} \right), \\
&\hat{h}_{V} = - \boldsymbol{V} \cdot \hat{ \boldsymbol{\sigma} }, 
\hspace{10pt}
\boldsymbol{V} = V_{x}\boldsymbol{e}_1 + V_{y}\boldsymbol{e}_2,
\label{eq:abvhm}
\end{align}
where $m$, $\mu$, $\alpha$ and $\beta$ denotes
the effective mass of an electron,
the chemical potential, the strength of the Rashba spin-orbit coupling
and the strength of the Dresselhaus[100] spin-orbit coupling,
respectively. The unit vectors in spin space is denoted by $\boldsymbol{e}_j$ 
with $j=1-3$.
The part of the Hamiltonian $\hat{h}_{V}$ denotes the Zeeman potential
induced by the in-plane external magnetic field.
When we define
\begin{align}
&p_{\pm} = \frac{1}{\sqrt{2}} \left( p_{x} \pm p_{y} \right), \\
&x_{\pm} = \frac{1}{\sqrt{2}} \left( x \pm y \right),
\end{align}
the commutation relations among
$p_{\pm}$ and $x_{\pm}$ become
\begin{align}
&\left[ x_{\nu},p_{\nu} \right] = i \hbar \delta_{\nu,\nu}, \\
&\left[ x_{\nu},x_{\nu} \right] = \left[ p_{\nu},p_{\nu} \right] = 0.
\end{align}
In this basis, the Hamiltonian is represented as
\begin{align}
\hat{h}_{RD}=\hat{h}_{{\rm kin}}^{'} + \hat{h}_{+} + \hat{h}_{-} + \hat{h}_{V}, 
\end{align}
where
\begin{align}
&\hat{h}_{{\rm kin}}^{'} =\frac{1}{2m} \left( p_{+}^{2} + p_{-}^{2} \right)
 - \mu,\\
&\hat{h}_{\pm} =  \lambda_{\pm} p_{\pm} \hat{\sigma}_{\pm}, \\
&\lambda_{\pm} = \frac{1}{\hbar} \left( \alpha \pm \beta \right),
\hspace{10pt}
\hat{\sigma}_{\pm} = \frac{1}{\sqrt{2}}\left( \hat{\sigma}_{2} \pm \hat{\sigma}_{1} \right).
\end{align}
As shown in Fig.~\ref{fig:appc}(a),
the direction $\boldsymbol{e}_+= (\boldsymbol{e}_1 + \boldsymbol{e}_2)/\sqrt{2}$ 
and $\boldsymbol{e}_-= (\boldsymbol{e}_1 - \boldsymbol{e}_2)/\sqrt{2}$
are orthogonal to each other in spin space.
When the strength of the Rashba spin-orbit coupling $\alpha$
and the strength of the Dresslhaus spin-orbit coupling $\beta$
are equal to each other (i.e., $\alpha = \beta$),
$\lambda_{-}$ becomes zero. Such electron systems have been studied 
in spintronics because they show unusual spin property so called spin-helix\cite{pstsp}. 
In addition,
we tune the Zeeman potential as $V_{x}=-V_{y}=V/\sqrt{2}$.
As shown in Fig.~\ref{fig:appc}(b),
the Hamiltonian is constructed by the two orthogonal components in the spin space,
\begin{align}
\hat{h}_{RD}=\hat{h}_{{\rm kin}}^{'} + \lambda_{+} p_{+} \hat{\sigma}_{+} + V \hat{\sigma}_{-}.
\end{align}
By multiplying a unitary matrix,
\begin{align}
\hat{d}= \frac{1}{\sqrt{2}}
\left[
\begin{array}{cc}
e^{-i \pi/8} & i e^{i \pi/8} \\
i e^{-i \pi/8} & e^{i \pi/8} \\
\end{array}
\right],
\end{align}
the Hamiltonian is transformed into
\begin{align}
\hat{h}_{RD}^{'} &= \hat{d}\; \hat{h}_{RD}\; \hat{d}^{\dagger} \nonumber\\
&=\hat{h}_{{\rm kin}}^{'} - \lambda_{+} p_{+} \hat{\sigma}_{3} - V \hat{\sigma}_{1}.
\end{align}
The Hamiltonian $\hat{h}_{RD}^{'}$ is equivalent to
the Hamiltonian with the Dresselhaus [110] spin-orbit coupling
and with the in-plane magnetic field.

Next, we introduce the proximity-induced $s$-wave pair potential.
The BdG Hamiltonian for the original basis in Eq.~(\ref{eq:abvhm}) is given by
\begin{align}
\check{H}_{RD} = \left[
\begin{array}{cc}
\hat{h}_{RD} & i \Delta_{0} \hat{\sigma}_{2} \\
- i \Delta_{0} \hat{\sigma}_{0} & - \hat{h}_{RD}^{*} \\
\end{array}
\right].
\end{align}
When $\alpha=\beta$ and
$V_{x}=-V_{y}=V/\sqrt{2}$, the BdG Hamiltonian is 
transformed into 
\begin{align}
\check{H}_{RD}^{'} &= \hat{D} \hat{H}_{RD} \hat{D}^{\dagger} \nonumber\\
&= \left[
\begin{array}{cc}
\hat{h}_{RD}^{'} & i \Delta_{0} \hat{\sigma}_{2} \\
- i \Delta_{0} \hat{\sigma}_{0} & - \hat{h}_{RD}^{'*} \\
\end{array}
\right],\\
\check{D} &= \left[
\begin{array}{cc}
\hat{d} & 0 \\
0 & \hat{d}^{*} \\
\end{array}
\right].
\end{align}
The BdG Hamiltonian $\check{H}_{RD}^{'}$ is unitary equivalent to
the BdG Hamiltonian in Eq.~(\ref{eq:original}).
Therefore, by applying the unitary transformations introduced in Appendix~\ref{sec:trbdg},
we can obtain the BdG Hamiltonian of the effective $p_{x}$-wave super conductor in
Eq. (\ref{eq:bdganap}).
As shown in Fig. \ref{fig:appc}(c),
we have to prepare the nanowire along the $x_{+}$ direction
and apply the magnetic field in the $x_{-}$ direction.
More generally, by applying the appropriate unitary transformation,
we also obtain the Hamiltonian $\check{H}_{RD}^{'}$
with
\begin{align}
\boldsymbol{V} =  V\left[ \frac{\sin\phi}{\sqrt{2}}\boldsymbol{e}_1 
- \frac{\sin\phi}{\sqrt{2}}\boldsymbol{e}_2 + \cos\phi \boldsymbol{e}_3 \right],
\end{align}
where $\phi$ is the arbitrary angle of the magnetic field.
Alternatively, it is possible to reach 
BdG Hamiltonian in Eq.~(\ref{eq:original})
when we 
prepare the nanowire along the $x_{-}$ direction
with $\alpha=-\beta$ and 
\begin{align}
\boldsymbol{V} =  V\left[ \frac{\sin\phi}{\sqrt{2}}\boldsymbol{e}_1 
+ \frac{\sin\phi}{\sqrt{2}}\boldsymbol{e}_2 + \cos\phi \boldsymbol{e}_3 \right].
\end{align}

\begin{thebibliography}{32}
\bibitem{maj1} E. Majorana, Nuovo Cimento, {\bf 14}, 171 (1937).
\bibitem{maj2} F.Wiczek, Nature Phys. {\bf 5}, 614 (2009)
\bibitem{tqc1} D. A. Ivanov, Phys. Rev. Lett. {\bf 86}, 268 (2001).
\bibitem{tqc2} J. D. Sau, D. J. Clarke, and S. Tewari, Phys. Rev B {\bf 84}, 094505 (2011).
\bibitem{psc1} N. Read and D. Green, Phys. Rev. B {\bf 61}, 10267 (2000).
\bibitem{psc2} A. Y. Kitaev, Phys. Usp. {\bf 44}, 131 (2001).
\bibitem{tisw} L. Fu and C. L. Kane, Phys. Rev. Lett. {\bf 100}, 096407 (2008).
\bibitem{smsc1} M. Sato, Y. Takahashi, and S. Fujimoto, Phys. Rev. Lett. {\bf 103}, 020401 (2009).
\bibitem{smsc2} J. D. Sau, R. M. Lutchyn, S. Tewari, and S. DasSarma, Phys. Rev. Lett. {\bf 104}, 040502 (2010).
\bibitem{smsc3} J. Alicea, Phys. Rev. B {\bf 81}, 125318 (2010).
\bibitem{smsc4} R. M. Lutchyn, J. D. Sau, and S. DasSarma, Phys. Rev. Lett. {\bf 105}, 077001 (2010).
\bibitem{smsc5} Y. Oreg, G. Refael, and F. von Oppen, Phys. Rev. Lett. {\bf 105}, 177002 (2010).
\bibitem{smsc6} T. D. Stanescu, R. M. Lutchyn, and S. DasSarma, Phys. Rev. B {\bf 84}, 144522 (2011).
\bibitem{hlsc} M. Sato and S. Fujimoto, Phys. Rev. B {\bf 79}, 094504 (2009).
\bibitem{scti} S. Sasaki, M. Kriener, K. Segawa, K. Yada, Y. Tanaka, M. Sato, and Y. Ando,
Phys. Rev. Lett. {\bf 107}, 217001 (2011).
\bibitem{mjsp1} A. C. Potter and P. A. Lee, Phys. Rev. Lett. {\bf 105}, 227003 (2010);
A. C. Potter and P. A. Lee, Phys. Rev. B {\bf 83}, 094525 (2011).
\bibitem{mjsp2} R. M. Lutchyn, T. D. Stanescu, and S. DasSarma, Phys. Rev. Lett. {\bf 106}, 127001 (2011).
\bibitem{mjsp3} M. Gibertini, F. Taddei, M. Polini, and R. Fazio, Phys. Rev. B {\bf 85}, 144525 (2012).
\bibitem{drso} G. Dresselhaus, Phys. Rev. {\bf 100}, 580-586 (1955).
\bibitem{flds} J. You, C. H. Oh, and V. Vedral, Phys. Rev. B {\bf 87}, 054501 (2013).
\bibitem{leggett} A.\ J.\ Leggett, Rev. Mod. Phys. \textbf{47}, 331 (1975).
\bibitem{buchholtz} L.~J.~Buchholtz and G.~Zwicknagl, Phys. Rev. B
\textbf{23}, 5788 (1981).
\bibitem{hara} J.~Hara and K.~Nagai, Prog. Theor. Phys. \textbf{74},
1237 (1986).
\bibitem{hu} C.~R.~Hu, Phys. Rev. Lett. \textbf{72}, 1526 (1994).
\bibitem{tanaka95} Y.\ Tanaka and S.\ Kashiwaya, Phys. Rev. Lett.
\textbf{74}, 3451 (1995).
\bibitem{chral} M. Sato, Y. Tanaka, K. Yada, and T. Yokoyama,
Phys. Rev. B {\bf 83}, 224511(2011).
\bibitem{yt05r} Y.\ Tanaka, Y.\ Asano, A.~A.\ Golubov, and S.\ Kashiwaya,
Phys.\ Rev.~B \textbf{72}, 140503(R) (2005).
\bibitem{yt04} Y.\ Tanaka and S.\ Kashiwaya, Phys.\ Rev.~B \textbf{70},
012507 (2004).
\bibitem{ya06} Y.\ Asano, Y.\ Tanaka, and S.\ Kashiwaya, Phys.\ Rev.\
Lett.\ \textbf{96}, 097007 (2006).
\bibitem{ya07} Y.\ Asano, Y.\ Tanaka, A.~A.\ Golubov, and S. Kashiwaya,
Phys.\ Rev.\ Lett.\ \textbf{99}, 067005 (2007).
\bibitem{ya11} Y.\ Asano, A.\ A.\ Golubov, Ya.\ V.\ Fominov, and Y.\
Tanaka, Phys. Rev. Lett. \textbf{107}, 087001 (2011).
\bibitem{dnps4} Y. Asano and Y. Tanaka, Phys. Rev B {\bf 87}, 104513 (2013).
\bibitem{dnps1} Y. Tanaka, S. Kashiwaya, and T. Yokoyama, Phys. Rev. B {\bf 71}, 094513 (2005).
\bibitem{grf1} P. A. Lee and D. S. Fisher, Phys. Rev. Lett. {\bf 47}, 882-885 (1981).
\bibitem{grf2} T. Ando, Phys. Rev. B {\bf 44}, 8017-8027 (1991).
\bibitem{btkf} G. E. Blonder, M. Tinkham, and T. M. Klapwijk, Phys .Rev. B {\bf 25}, 4515 (1982).
\bibitem{jsphs} Y. Asano, Phys. Rev. B {\bf 63}, 052512 (2001).
\bibitem{likharev} K.\ K.\ Likharev, Rev. Mod. Phys. \textbf{51}, 101
(1979).
\bibitem{fwts} L. L. Foldy and S. S. Wouthuysen, Phys. Rev. {\bf 78}, 29-36 (1950).
\bibitem{pstsp} B. A. Bernevig, J. Orenstein and S. C. Zhang,
Phys. Rev. Lett. {\bf 97}, 236601 (2006).


\end{thebibliography}

\end{document}